%% file: main.tex
\documentclass[twocolumn,trackchanges,floatfix]{aastex631}
\usepackage{amsmath}
\usepackage{longtable}
\usepackage{multirow}
\usepackage{booktabs}
\usepackage[english]{babel}
\usepackage{url}
\usepackage{natbib}
\usepackage{gensymb}
\usepackage{subfigure}
\DeclareUnicodeCharacter{2212}{-}
\usepackage{amsmath,bm}
\usepackage{longtable}
\usepackage{multirow}
\usepackage{threeparttable}
\usepackage{times}
\usepackage{enumitem}
\usepackage{lipsum}[1-2]
\usepackage{array}

\newcommand{\GC}{\text{GC}}

\newcommand{\kms}{km s$^{-1}$}
\newcommand{\zmax}{$Z_{\rm max}$}
\newcommand{\Rmax}{$R_{\rm max}$}

\shorttitle{Candidate Stars in the VMP/EMP Galactic Disk System}
\shortauthors{Hong et al.}
\begin{document}

\title{Candidate Members of the VMP/EMP Disk System of the Galaxy from the SkyMapper and SAGES Surveys}

\author[0000-0002-2453-0853]{Jihye Hong}
\affiliation{Department of Physics and Astronomy, University of Notre Dame, Notre Dame, IN 46556, USA}
\affiliation{Joint Institute for Nuclear Astrophysics -- Center for the Evolution of the Elements (JINA-CEE), USA}

\author[0000-0003-4573-6233]{Timothy C. Beers}
\affiliation{Department of Physics and Astronomy, University of Notre Dame, Notre Dame, IN 46556, USA}
\affiliation{Joint Institute for Nuclear Astrophysics -- Center for the Evolution of the Elements (JINA-CEE), USA}

\author[0000-0001-5297-4518]{Young Sun Lee}
\affiliation{Department of Astronomy and Space Science, Chungnam National University, Daejeon 34134, Republic of Korea}
\affiliation{Department of Physics and Astronomy, University of Notre Dame, Notre Dame, IN 46556, USA}

\author[0000-0003-3250-2876]{Yang Huang}
\affiliation{School of Astronomy and Space Science, University of Chinese Academy of Sciences, Beijing 100049, People's Republic of China}
\affiliation{CAS Key Lab of Optical Astronomy, National Astronomical Observatories, Chinese Academy of Sciences, \\ Beijing 100101, People's Republic of China}

\author[0000-0002-5661-033X]{Yutaka Hirai}
\altaffiliation{JSPS Research Fellow}
\affiliation{Department of Physics and Astronomy, University of Notre Dame, Notre Dame, IN 46556, USA}
\affiliation{Astronomical Institute, Tohoku University, 6-3 Aoba, Aramaki, Aoba-ku, Sendai, Miyagi 980-8578, Japan}
\affiliation{Joint Institute for Nuclear Astrophysics -- Center for the Evolution of the Elements (JINA-CEE), USA}

\author[0009-0006-7257-913X]{Jonathan Cabrera Garcia}
\affiliation{Department of Physics and Astronomy, University of Notre Dame, Notre Dame, IN 46556, USA}
\affiliation{Joint Institute for Nuclear Astrophysics -- Center for the Evolution of the Elements (JINA-CEE), USA}

\author[0000-0001-9723-6121]{Derek Shank}
\affiliation{Department of Physics and Astronomy, University of Notre Dame, Notre Dame, IN 46556, USA}
\affiliation{Joint Institute for Nuclear Astrophysics -- Center for the Evolution of the Elements (JINA-CEE), USA}

\author[0000-0003-3535-504X]{Shuai Xu}
\affiliation{Institute for Frontiers in Astronomy and Astrophysics, Beijing Normal University, Beijing 102206, China}
\affiliation{Department of Astronomy, Beijing Normal University, Beijing, 100875, P.R.China}

\author[0000-0003-2471-2363]{Haibo Yuan}
\affiliation{Institute for Frontiers in Astronomy and Astrophysics, Beijing Normal University, Beijing 102206, China}
\affiliation{Department of Astronomy, Beijing Normal University, Beijing, 100875, P.R.China}

\author[0000-0001-9178-3992]{Mohammad K.\ Mardini}
\affiliation{Department of Physics, Zarqa University, Zarqa 13110, Jordan}
\affiliation{Jordanian Astronomical Virtual Observatory, Zarqa University, Zarqa 13110, Jordan}

\author[0000-0001-9012-8320]{Thomas Catapano}
\affiliation{Department of Physics and Astronomy, University of Notre Dame, Notre Dame, IN 46556, USA}
\affiliation{Joint Institute for Nuclear Astrophysics -- Center for the Evolution of the Elements (JINA-CEE), USA}

\author[0000-0002-8980-945X]{Gang Zhao}
\affiliation{CAS Key Lab of Optical Astronomy, National Astronomical Observatories, Chinese Academy of Sciences, \\ Beijing 100101, People's Republic of China}

\author[0000-0002-6790-2397]{Zhou Fan}
\affiliation{CAS Key Lab of Optical Astronomy, National Astronomical Observatories, Chinese Academy of Sciences, \\ Beijing 100101, People's Republic of China}

\author[0000-0001-6637-6973]{Jie Zheng}
\affiliation{CAS Key Lab of Optical Astronomy, National Astronomical Observatories, Chinese Academy of Sciences, \\ Beijing 100101, People's Republic of China}

\author[0000-0002-9702-4441]{Wei Wang}
\affiliation{CAS Key Lab of Optical Astronomy, National Astronomical Observatories, Chinese Academy of Sciences, \\ Beijing 100101, People's Republic of China}

\author[0000-0003-0173-6397]{Kefeng Tan}
\affiliation{CAS Key Lab of Optical Astronomy, National Astronomical Observatories, Chinese Academy of Sciences, \\ Beijing 100101, People's Republic of China}

\author[0000-0003-2868-8276]{Jingkun Zhao}
\affiliation{CAS Key Lab of Optical Astronomy, National Astronomical Observatories, Chinese Academy of Sciences, \\ Beijing 100101, People's Republic of China}

\author[0009-0000-4835-7525]{Chun Li}
\affiliation{CAS Key Lab of Optical Astronomy, National Astronomical Observatories, Chinese Academy of Sciences, \\ Beijing 100101, People's Republic of China}

\accepted{Apr18}

\begin{abstract}

Photometric stellar surveys now cover a large fraction of the sky, probe to fainter magnitudes than large-scale spectroscopic surveys, and are relatively free from the target-selection biases often associated with such studies. Photometric-metallicity estimates that include narrow/medium-band filters can achieve comparable accuracy and precision to existing low-resolution spectroscopic surveys such as SDSS/SEGUE and LAMOST. Here we report on an effort to identify likely members of the Galactic disk system among the very metal-poor (VMP; [Fe/H] $\leq$ --2) and extremely metal-poor (EMP; [Fe/H] $\leq$ --3) stars. Our analysis is based on an initial sample of $\sim11.5$ million stars with full space motions selected from the SkyMapper Southern Survey (SMSS) and Stellar Abundance and Galactic Evolution Survey (SAGES). After applying a number of quality cuts to obtain the best available metallicity and dynamical estimates, we analyze a total of 
$\sim$5.86 million stars in the combined SMSS/SAGES sample. We employ two techniques that, depending on the method, identify between 876 and 1,476 VMP stars (6.9\%-11.7\% of all VMP stars) and between 40 and 59 EMP stars (12.4\%-18.3\% of all EMP stars) that appear to be members of the Galactic disk system on highly prograde orbits (v$_{\phi} > 150$ \kms). The total number of candidate VMP/EMP disk-like stars is 1,496, the majority of which have low orbital eccentricities, ecc $\le 0.4$; many have ecc $\le 0.2$. The large fractions of VMP/EMP stars associated with the Milky Way disk system strongly suggest the presence of an early forming ``primordial" disk.

\end{abstract}


\section{Introduction}\label{sec:introduction}

Over the past few decades, large-scale spectroscopic surveys, such as the HK Survey \citep{Beers1985, Beers1992}, the Hamburg/ESO Survey \citep{Christlieb2003}, the Sloan Digital Sky Survey \citep[SDSS;][]{York2000}, SEGUE \citep{Yanny2009, Rockosi2022}, RAVE \citep{Steinmetz2006}, LAMOST \citep{Deng2012, Zhao2012}, GALAH \citep{DeSilva2015}, APOGEE \citep{Majewski2017}, the H3 Survey \citep{Conroy2019}, the $Gaia$-ESO survey \citep{gilmore2022}, and the $Gaia$ mission \citep{GaiaCollaboration2023b} have changed the paradigm of observational studies by providing detailed chemical and kinematic information for numerous stars in the Milky Way (MW), in particular for the relatively rare very metal-poor (VMP; [Fe/H] $\leq$ --2) and extremely metal-poor (EMP; [Fe/H] $\leq$ --3) stars. 

In a series of recent papers, \citet{An2020, An2021a, An2021b} and \citet{An2023} have constructed ``blueprints" of the MW's stellar populations from analyses of the orbital rotation (inferred from proper motions and distance estimates alone) as a function of carefully calibrated photometric-metallicity estimates for stars with available broadband $ugriz$ from SDSS/SEGUE and other surveys. This approach has proven quite powerful. Among other results, these authors not only identified previously known substructures and confirmed the presence of the inner- and outer-halo populations but also demonstrated that the metal weak-thick disk \citep[MWTD;][]{Norris1985,carollo2007,Carollo2010,Beers2014} is a separable population with lower metallicity and rotation that lags the canonical thick disk, as shown in \citet{Carollo2019}. In addition, they identified a continuous sequence of stars in the rotational velocity vs. metallicity space that may be associated with a starburst event when the earlier disk system encountered $Gaia$-Sausage-Enceladus \citep[GSE;][]{Belokurov2018,Haywood2018,Helmi2018}. Evidence for this starburst event is also reported in \citet{Lee2023}.

Whether surveys to identify likely metal-poor (MP) stars are performed with  
fiber-fed spectrographs such as SDSS or LAMOST or broadband photometric efforts such as SEGUE or Pan-STARRS \citep{Chambers2016}, it is challenging to avoid target selection biases that can confound the relative contributions of stars with different metallicity to the recognized Galactic components. In addition, the first step in surveys dedicated to finding low-metallicity stars is often to limit the regions of the sky under consideration to higher Galactic latitude (e.g., $|b| > 30^{\circ}$), precluding identification of substantial numbers of VMP/EMP stars in the disk system of the MW.

Nonetheless, recent papers have provided identifications of VMP/EMP (and a handful of ultra MP, UMP; [Fe/H] $\leq -4$) stars in the MW with disk-like orbits, based on medium-resolution and, and in some cases, high-resolution, spectroscopic follow-up (see, e.g., \citealt{Schlaufman2018, Sestito2019, Sestito2020}; \citealt{DiMatteo2020}; \citealt{Venn2020}; \citealt{Carter2021}; \citealt{Cordoni2021}; \citealt{Fernandez-Alvar2021}; \citealt{Limberg2021}, \citealt{Mardini2022a, Mardini2022b, Mardini2024}; \citealt{Carollo2023}, and references therein).

Over the past few years, photometric surveys based on combinations of narrow-band and medium- to broad-band filters have been (or are being) executed (e.g., SkyMapper; \citealt{Keller2007}, the Pristine Survey; \citealt{Starkenburg2017}, Stellar Abundance and Galactic Evolution Survey (SAGES); \citealt{Zheng2018}, J-PLUS; \citealt{Cenarro2019}, and S-PLUS; \citealt{MendesdeOliveira2019}). Typically, such surveys do not avoid regions of the MW at lower Galactic latitudes, other than those limited by very high interstellar extinction and reddening or crowding. As a result, VMP/EMP stars in the MW's disk system have been increasingly discovered, though their numbers are still relatively small.

The SkyMapper Southern Survey Data Release 2 \citep[SMSS DR2;][]{Onken2019} was carefully recalibrated by \citet{Huang2021a}, and used by \citet{Huang2022} to derive stellar parameters, luminosity classifications, and metallicity estimates for over 24 million stars in the Southern Hemisphere. These authors derived effective temperatures ($T_{\rm eff}$) by adopting metallicity-dependent  $T_{\rm eff}$-color relations constructed from Gaia $(G_{\rm BP}-G_{\rm RP})_0$, LAMOST $T_{\rm eff}$, and [Fe/H]. The effective temperature scale of LAMOST has been shown to agree with that of direct measurements \citep{Huang2015b}. They adopted Bayesian distance estimates \citep{Bailer-Jones2021} and ages from the PAdova and tRieste Stellar Evolutionary Code \citep[\texttt{PARSEC};][]{Bressan2012,Marigo2017} isochrones.
In addition, an
empirical metallicity-dependent stellar-locus method \citep{Yuan2015} was used to estimate the photometric metallicity, 
with combinations of the SMSS narrow/medium $u$- and $v$-band filter magnitudes, the $G_{\rm BP}$ magnitude from the $Gaia$ ultra wide band prism spectra, and a maximum likelihood approach \citep{Huang2022}.

The recently completed SAGES \citep{Fan2023}, which employs similar, but not identical, filters to SMSS, has been employed by \citet{Huang2023} to obtain stellar parameters, luminosity classifications, and metallicity estimates for nearly 26 million stars in the Northern Hemisphere. 

Here we identify 1,496 VMP and 61 EMP candidate stars with disk-like orbits populating the rapidly rotating disk system of the MW (v$_{\phi} > $ 150 \kms), selected from a subset of roughly 11.5 million stars from the SMSS and SAGES photometric surveys with available radial velocities (RVs), proper motions, and other astrometric data from which full space motions are derived. We approximately separate stars with disk-like orbits from stars with halo-like orbits by two criteria that have been commonly used in the literature \citep{Haywood2018,DiMatteo2020,Mardini2022a,Bellazzini2024}, and then consider their relative fractions at low metallicities. 

This paper is organized as follows: In Section~\ref{sec:dataandmethods}, we describe the data sets we employ and the choices made for the adopted metallicity estimates, as well as the derivation of dynamical parameters. In this section we also describe two methods that have been commonly used to identify stars with potential disk-like orbits. In Section~\ref{sec:results}, we present maps of the orbital rotational velocities of the stars as a function of [Fe/H], where potential VMP/EMP candidates with disk-like orbits can already be seen, and compare their relative fractions as a function of [Fe/H]. In Section~\ref{sec:discussion}, we present a discussion, along with conjectures on the origins of VMP/EMP disk-like stars based on interpretations from numerical simulations. We conclude with a summary and future prospects in Section~\ref{sec:summary}.

\section{Data and Methods}\label{sec:dataandmethods}

\subsection{Data}\label{sec:data}

\citet{Huang2022} derived stellar parameters, including metallicity estimates, for more than 19 million dwarfs and 5 million giants over essentially the entire Southern Hemisphere from SMSS DR2, including about 731,000 VMP and 27,000 EMP stars. If we restrict their sample to stars with available RVs from $Gaia$ DR3 \citep{GaiaCollaboration2023b} and other sources, the number of stars is about 7.4 million, including roughly 56,000 VMP and 2,300 EMP stars.

SAGES observed slightly less than half of the Northern Hemisphere. Notably, SAGES did not cover a large fraction of the north galactic pole (NGP), while SMSS covered the entire south galactic pole (SGP). Another crucial difference between SAGES and SMSS is that the central wavelength of the SAGES $v$-band filter is shifted redward relative to the SMSS $v$-band filter by about $110~\text{\AA}$, so it fully includes the region of the Ca II H \& K lines \citep{Zheng2018}. \citet{Huang2023} used a similar approach to \citet{Huang2022}, and obtained effective temperatures, luminosity classifications based on surface gravity, and metallicity estimates for over 26 million stars, including some 874,000 VMP and 13,000 EMP stars from SAGES DR1 \citep{Fan2023}. About 4.1 million stars in this catalog have available RVs, including roughly 41,000 VMP and 1,900 EMP stars.

For this study, we begin with a sample of about 7.4 million stars from SMSS and 4.1 million stars from SAGES with available RVs, proper motions, and distance estimates, as provided in the catalogs from \citet{Huang2022} and \citet{Huang2023}, respectively. After combining these data sets, binary stars photometrically classified by \citet{Huang2022, Huang2023} and cool dwarfs ($T_\texttt{\rm eff} <$ 4,500 K) have been removed. We have additionally applied a more restrictive cut on the \texttt{bp\_rp\_excess\_factor},  $<$ 0.12 $\times$ (BP-RP)$_{0}$ + 1.13/1.14 cuts for dwarfs/giants respectively, as in \citet{Xu2022}, and on the renormalized unit weight error (RUWE) $<$ 1.1, in order to exclude possible binary stars.
Cuts based on an empirical isochrone, similar to the \texttt{PARSEC} \citep{Bressan2012,Marigo2017} isochrone with [Fe/H] = $-2$ at age = 12 Gyr, were also applied to eliminate the significant contamination from metal-rich stars that could masquerade as VMP/EMP stars. These restrictions removed a total of about 3.3 million stars.

Moreover, we removed about 940 likely stellar globular cluster members based on the catalogs of \citet{Harris2010}\footnote{\url{https://physics.mcmaster.ca/~harris/mwgc.dat}} and \citet{Baumgardt2021}. 
Finally, in order to diminish the effect of reddening on the derived metallicities (of particular importance for stars near the disk), we only included stars with $E(B-V) \leq$ 0.3, excluding a total of about 17,000 stars. More discussion about the extinction cut is provided in the Appendix.

Metallicity estimates for the stars in our sample are based on calibrated $u - G_{\rm BP}$ colors and $v - G_{\rm BP}$ colors, a combination of the $u/v$bands from SMSS/SAGES and the ultra wide-band $Gaia$ $G_{\rm BP}$ prism spectra (\citealt{Huang2022, Huang2023}). As has been noted previously, the colors involving the $u$ band have a greater sensitivity to the presence of enhanced carbon in a star than those involving the $v$ band. For this reason, and in order to provide the best available metallicities, we do not include stars for which only $u$ band metallicity estimates are available, those that have a difference between the $u$-band- and $v$-band-based abundances greater than 0.5\,dex, and stars with estimated metallicity errors greater than 0.5\,dex. See the Appendix for a justification and full discussion of the cuts that are applied to the sample prior to assigning final adopted metallicities. Note that we refer to the 
photometric-metallicity estimates as [Fe/H] in this study, unless otherwise indicated.

\subsection{Dynamical Parameters}\label{sec:dyn_params}

Orbital parameters for the stars in our combined sample are determined using their 6D astrometric parameters (positions, RVs, proper motions, and distance estimates from \citealt{Huang2022, Huang2023}), as well as their corresponding errors, as inputs to the Action-based GAlaxy Modelling Architecture\footnote{\url{http://github.com/GalacticDynamics-Oxford/Agama}} (\texttt{AGAMA}) package \citep{Vasiliev2019}, adopting the solar positions and peculiar motions described in \citet{Shank2022b}\footnote{We adopt a solar position of ($-8.249$, 0, 0) kpc (\citealt{GRAVITYCollaboration2020}) and solar peculiar motion ($U$, $V$ $W$), about the Local Standard of Rest (LSR) of (11.1,12.24,7.25) km $\text{s}^{-1}$ (\citealt{Schonrich2010}), where $V_\text{LSR} = 238.5 \, \text{km} \, \text{s}^{-1}$, defined as $V_{\text{LSR}} = V_{\odot} - V$ and $V_{\odot} = 250.70$ $\, \text{km} \, \text{s}^{-1}$, determined from \citet{Reid2020} based on our choice of solar position and using the proper motion of the center of the Galaxy (Srg A*) of $-6.411 \, \text{mas} \, \text{yr}^{-1}$.}, and the gravitational potential \texttt{MW2017} \citep{McMillan2017}.

Similar to \citet{Shank2022b}, we input quantities through the orbital integration process in \texttt{AGAMA} to calculate the cylindrical velocities (v$_r$, v$_{\phi}$, v$_z$), cylindrical actions (J$_{r}$, J$_{\phi}$, J$_{z}$), orbital specific energy ($E$), $r_{\text{apo}}$, $r_{\text{peri}}$, eccentricity (ecc), \zmax\ (the maximum orbital distances reached by stars from the Galactic plane), and \Rmax\ (the maximum apocentric distance projected onto the Galactic plane), along with their associated errors\footnote{Due to the very large number of stars in our sample, we estimated errors on \Rmax, unlike for the other dynamical parameters, by running the input quantities and their errors 50 times through \texttt{AGAMA} using a random sample of 10,000 stars, and assume that the relative errors apply to all stars.}. Stars that are possibly unbound ($E > 0 \, \text{km}^{2} \, \text{s}^{-2}$) were identified and removed. This resulted in a total of 10.8 million from the initial 11.5 million stars that are suitable for our kinematic analysis. For our present purpose, we only included stars having derived errors less than 25 \kms\ in their orbital rotation velocities and relative errors $\leq$ 30\% in \zmax\ and \Rmax, which removed about 165,000 stars from the combined sample. This produced a final sample of approximately 5.86 million stars, including 4.07 million SMSS and 1.79 million SAGES stars, which we refer to as the SMSS/SAGES sample hereafter.

\begin{figure*}
    \centering
    \includegraphics[width=\linewidth]{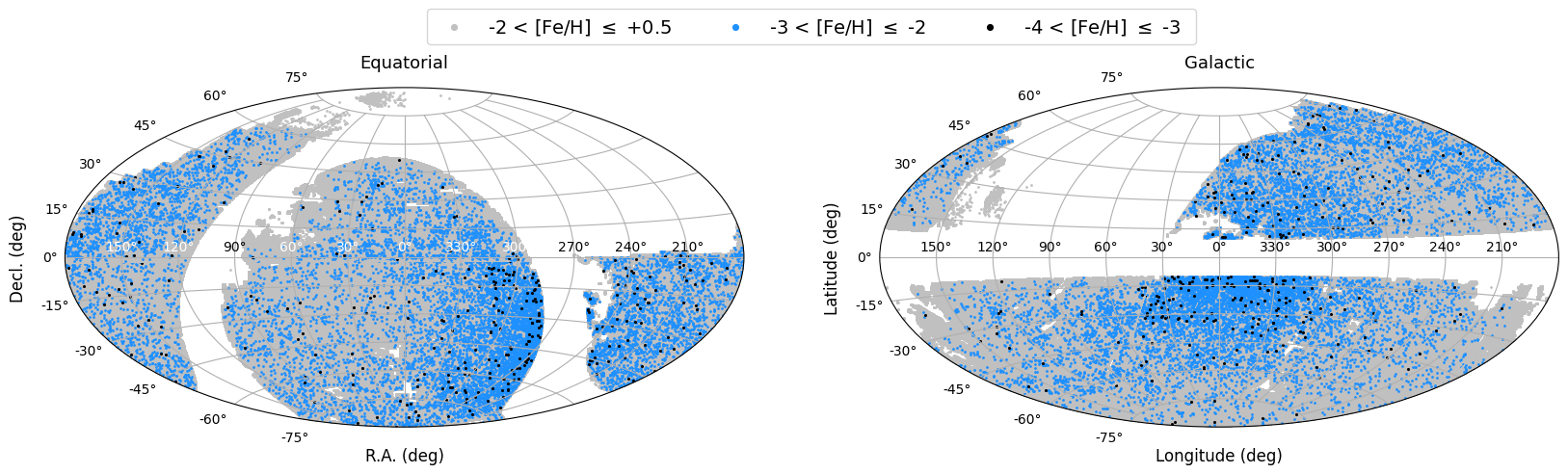}
    \caption{Mollweide projection of the positions for 5.86 million stars selected from the Southern Hemisphere (SMSS) and Northern Hemisphere (SAGES) photometric surveys, in equatorial (left panel) and Galactic (right panel) coordinates. The metallicity of this final sample is based on calibrated $v - G_{\rm BP}$ colors and a combination of the $u/v$ bands from SMSS/SAGES (\citealt{Huang2022, Huang2023}). 
    See the Appendix for a discussion of the cuts that are applied to the sample prior to assigning the final metallicities. The gray filled circles indicate the stars with derived metallicities in the range $-2 < $ [Fe/H] $\leq +0.5$, the light-blue filled circles are stars with $-3 < $ [Fe/H] $\leq -2$, and the black filled circles represent stars with $-4 < $ [Fe/H] $\leq -3$. The stars shown all have available RVs and astrometric information. For the purpose of our analysis, we exclude stars identified as likely binaries by \cite{Huang2022, Huang2023}, cool dwarfs ($T_\texttt{\rm eff} <$ 4,500\, K), metal-rich stars masquerading as VMP/EMP stars (see text), and likely members of recognized globular clusters. For stars in common between the two surveys, we have used the average value of the photometric-metallicity estimates.}
    \label{Fig:obsaarea}
\end{figure*}

Figure~\ref{Fig:obsaarea} shows the sky distribution in equatorial and Galactic coordinates for the final 5.86 million stars of the combined SMSS/SAGES sample. The gray filled circles indicate the stars with  $-2 < $ [Fe/H] $\leq +0.5$, the light-blue filled circles are stars with $-3 < $ [Fe/H] $\leq -2$, and the black filled circles represent stars with $-4 < $ [Fe/H] $\leq -3$. The $\sim 143,000$ stars in common between the two surveys have differences in [Fe/H] from the $v$ band with a median value of only 0.02\,dex; we have adopted an average of these determinations for these stars.

\begin{figure*}
    \centering
    \includegraphics[width=0.95\linewidth]{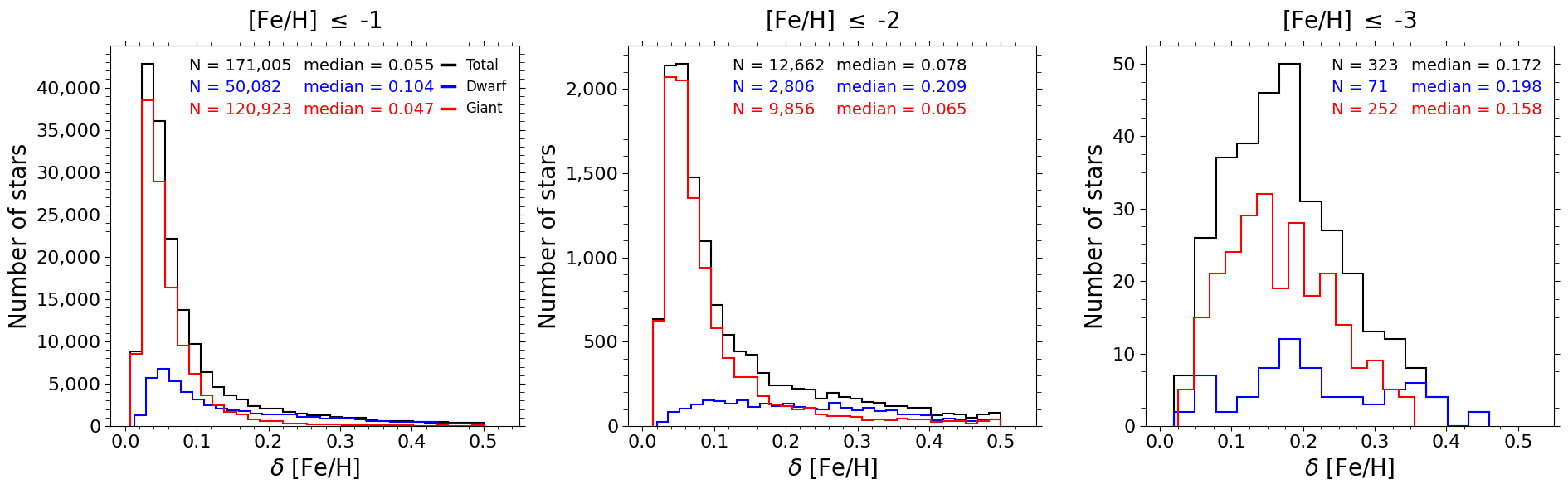}
    \caption{Histogram of the errors in photometric-metallicity estimates ($\delta$\,[Fe/H]) for the final SMSS/SAGES sample.  From left to right, the panels correspond to MP ([Fe/H] $\leq -1$), VMP ( [Fe/H] $\leq -2$), and EMP ([Fe/H] $\leq -3$) sub-samples, respectively. The black, blue, and red histograms represent the total, dwarf, and giant stars in each metallicity range, respectively. The number of stars and the median values of $\delta$\,[Fe/H] are indicated in the legend of each panel.}
    \label{Fig:feh_error}
\end{figure*}

Figure \ref{Fig:feh_error} shows the distribution of the errors in the photometric-metallicity estimates ($\delta$\,[Fe/H]) for the combined SMSS/SAGES sample. The left, middle, and right panels provide the results for the subsamples of stars with [Fe/H] $\leq -1$, $\leq -2$, and $\leq -3$, respectively. The legends in each panel indicate the median errors for all stars in the listed metallicity range, and the errors for stars classified as dwarfs and giants.  As can be seen, the errors increase with decreasing metallicity, as expected, but still remain reasonably low 
(median errors on the order of 0.1 to 0.2\,dex). Note that the external errors are somewhat larger, on the order of 0.25-0.35\,dex (see the discussion in the Appendix).

\subsection{Separation of Disk and Halo Stars}\label{sec:Separation} 

Previous analyses of the nature of stellar orbits in the MW have used a variety of techniques to separate stars on disk-like orbits from stars on halo-like orbits. Two simple approaches are described below.

\subsubsection{Maximum Height of Orbits}\label{sec:zmax}

This approach, employed by \citet{Beers2014}, \citet{Sestito2020, Sestito2021}, \citet{Limberg2021}, \citet{Mardini2022a}, and \citet{Bellazzini2024}, identifies stars in the disk-like and halo-like dynamical populations by assigning stars with \zmax\ $\leq 3$ kpc to disk-like orbits and those with \zmax\ $> 3$ kpc to halo-like orbits. Often, an additional criterion is adopted to identify the stars in the disk system by demanding that they be on highly prograde orbits. We follow a similar approach to that described below, with a further division of the stars on disk-like orbits into those with \zmax\ $\leq 1$ kpc, in an attempt to identify possible VMP/EMP thin-disk stars. 

\vskip 1cm
\subsubsection{``Wedges" in the Haywood Diagram}

Following \citet{Haywood2018}, we have also used plots of \zmax\ vs. 
arctan\,$(Z_{\rm max}{/}R_{\rm max})$, which redistributes our sample stars into discrete wedges, corresponding to different dynamical populations, a method also employed by \citet{Schuster2012}, \citet{DiMatteo2020}, \citet{Kim2021}, and \citet{Koppelman2021}.

Here $R_{\rm max}$ is defined as the projection of $r_{\text{apo}}$ onto the Galactic plane, via the simple geometric relationship \Rmax = $\sqrt{r_{\text{apo}}^2-Z_{\text{max}}^2}$.  Note that, for simplicity of notation, below we define an ``inclination angle" ($IA$) to represent arctan\,$(Z_{\rm max}{/}R_{\rm max})$.  It should be kept in mind that \zmax\ and \Rmax\ are derived from the full ensemble of orbits traced by a given star, so their $IA$ is representative of that complete set, not a single orbit or an average of the orbits.

\begin{figure}[t!]
\centering
\includegraphics[width=0.48\textwidth]{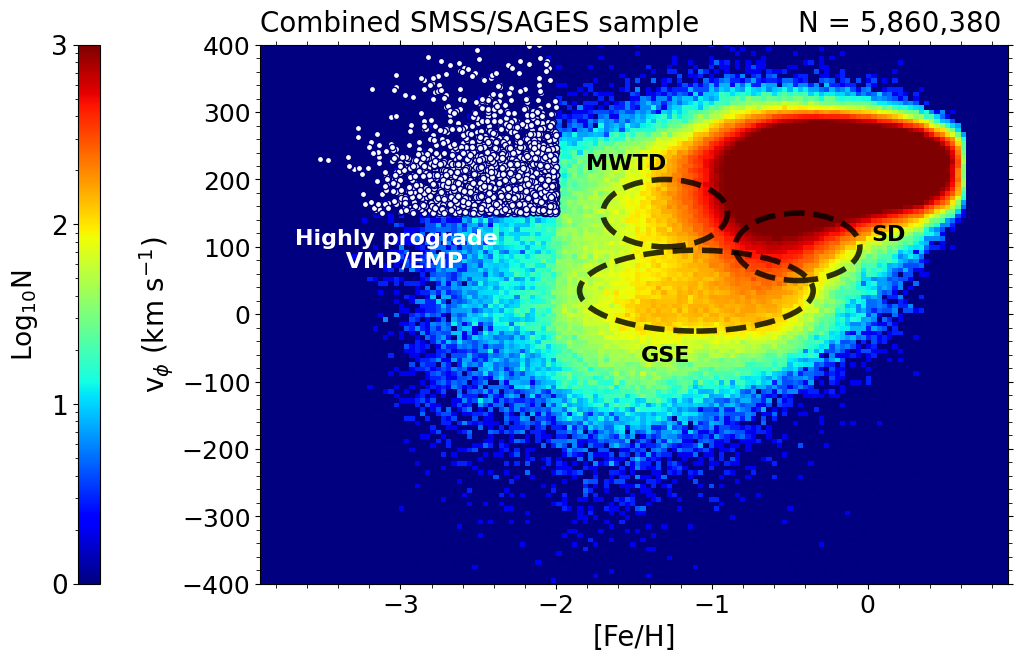}
\includegraphics[width=0.48\textwidth]{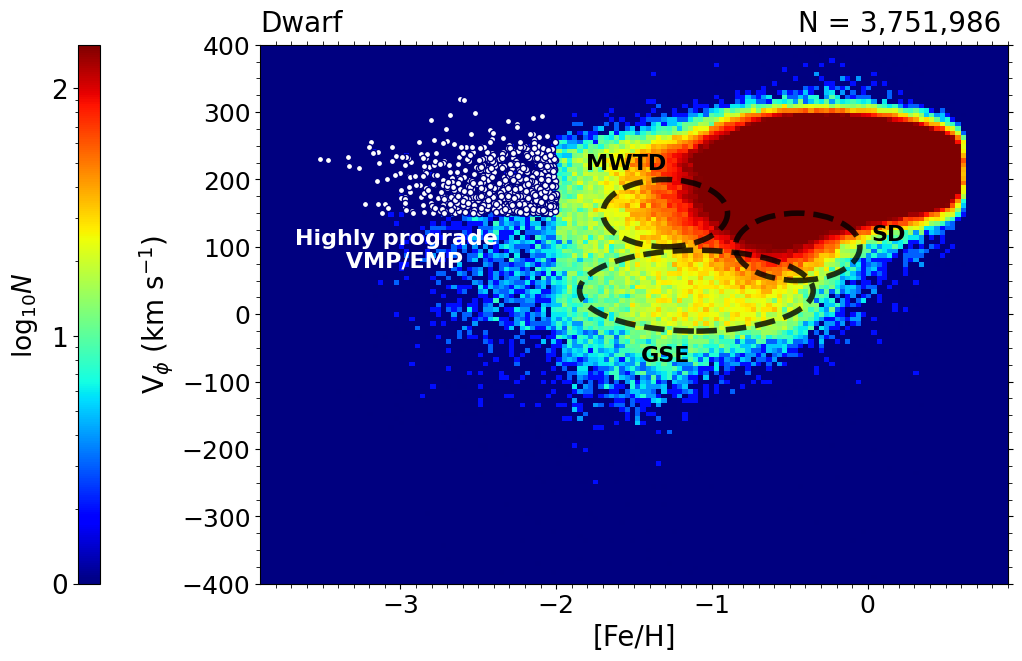}
\includegraphics[width=0.48\textwidth]{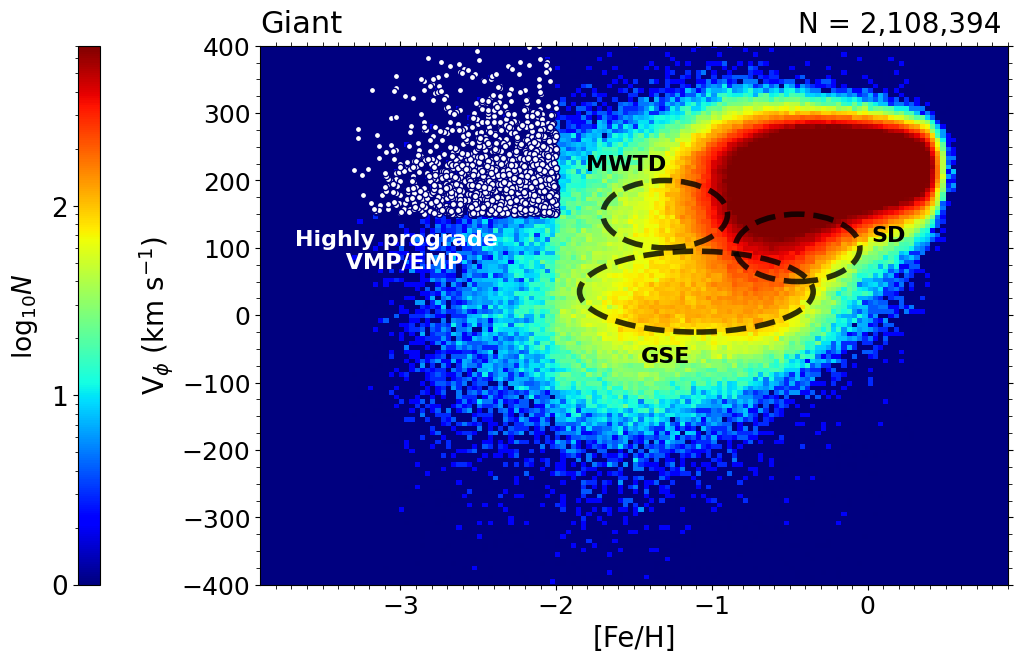}
\caption{Top panel: rotational velocity distribution (v$_{\phi}$) of the SMSS/SAGES sample as a function of photometric metallicity ([Fe/H]). The number density is color-coded on a logarithmic scale. The MWTD, the SD, and the GSE substructure are marked with black dashed ellipses. The highly prograde VMP/EMP candidates are shown by white circles. Middle panel: same as the top panel, but for dwarfs. Bottom panel: same as the top panel, but for giants. The number of VMP/EMP stars with v$_{\phi} > 150$ km s$^{-1}$ is about 2,150, including 650 dwarfs and 1,500 giants.}
    \label{Fig:fehvphi}
\end{figure}

\section{Results}\label{sec:results}

In this section, we identify about 12,700 VMP/EMP stars over the full range of the rotational velocities of the final 5.86 million stars in the combined SMSS/SAGES sample. Among these metal-deficient stars, we closely examine the 2,150 rapidly rotating VMP/EMP stars, in order to classify them as on halo-like or disk-like orbits.

\subsection{v$_{\phi}$ vs. \rm{[Fe/H]}}\label{sec:fehvphiresult}

Figure \ref{Fig:fehvphi} shows plots of stellar number density for our sample in the rotational velocity vs. photometric-metallicity plane. The top panel indicates the total combined SMSS/SAGES sample color-coded on a logarithmic scale. The rapidly rotating canonical disk system (comprising both the thin and thick disk) is most visible for [Fe/H] $> -1$. In addition, as reported in the series of papers by \citet{An2020, An2021a, An2021b}, the MWTD, the Splashed Disk (SD), and a hint of the GSE substructure can be seen in the black dashed ellipses. However, the VMP/EMP stars in the rapidly rotating disk region (v$_{\phi} > 150$ \kms; the average value of rotational velocity for the MWTD from \citealt{Carollo2010}), are less visible in the number density map than the other components. Thus, we represent the $\sim$2,150 highly prograde VMP/EMP candidates with white circles (out of a total number of about 12,700 VMP/EMP stars). The middle and bottom panels are the same v$_{\phi}$ vs. [Fe/H] plane, but for dwarfs and giants, respectively. There are about 2,800 total VMP/EMP dwarfs and 9,900 total VMP/EMP giants. The subsets of these stars with v$_{\phi} > 150$ \kms\ are roughly 650 VMP/EMP dwarfs and 1,500 VMP/EMP giants, respectively. We point out that, at this stage, we have not separated the VMP/EMP stars with disk-like orbits from those with halo-like orbits. However, from inspection of the middle and bottom panels, it is clear that the distribution of rotational velocity for the VMP/EMP giants (which are expected to contain a greater fraction of halo-like orbits) in the bottom panel stands in contrast to that of the VMP/EMP dwarfs, seen in the middle panel.

\begin{figure*}
    \includegraphics[width=0.345\textwidth]{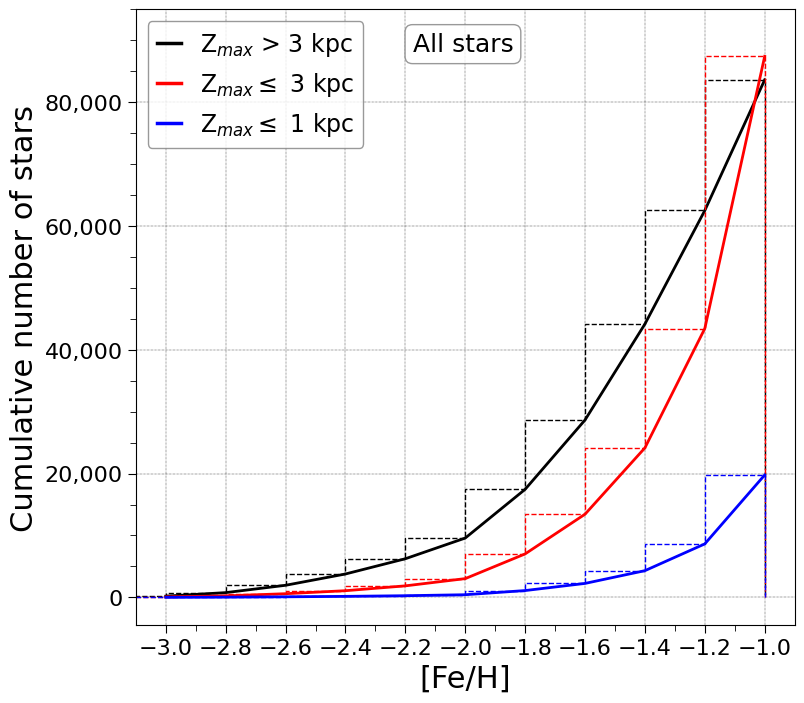}
    \includegraphics[width=0.325\textwidth]{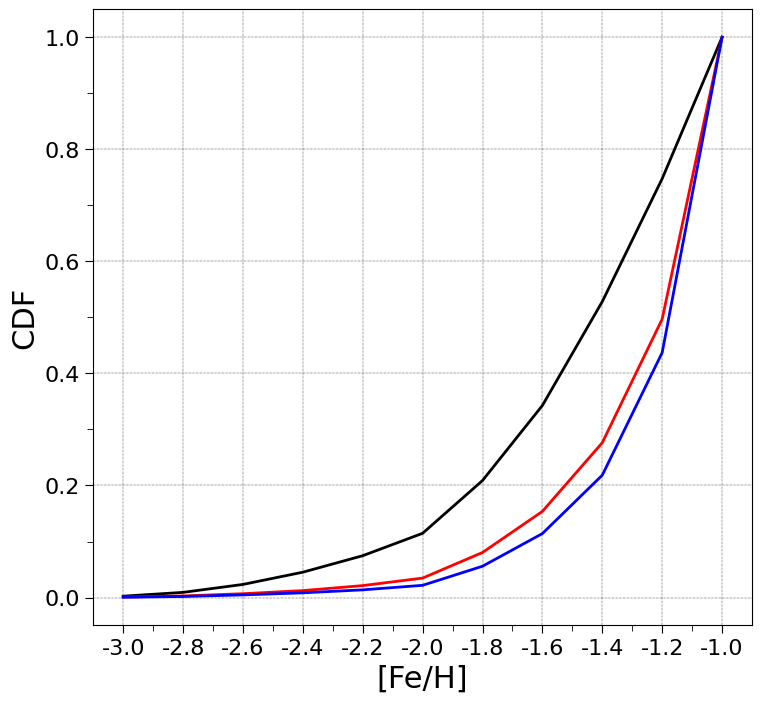}
    \includegraphics[width=0.325\textwidth]{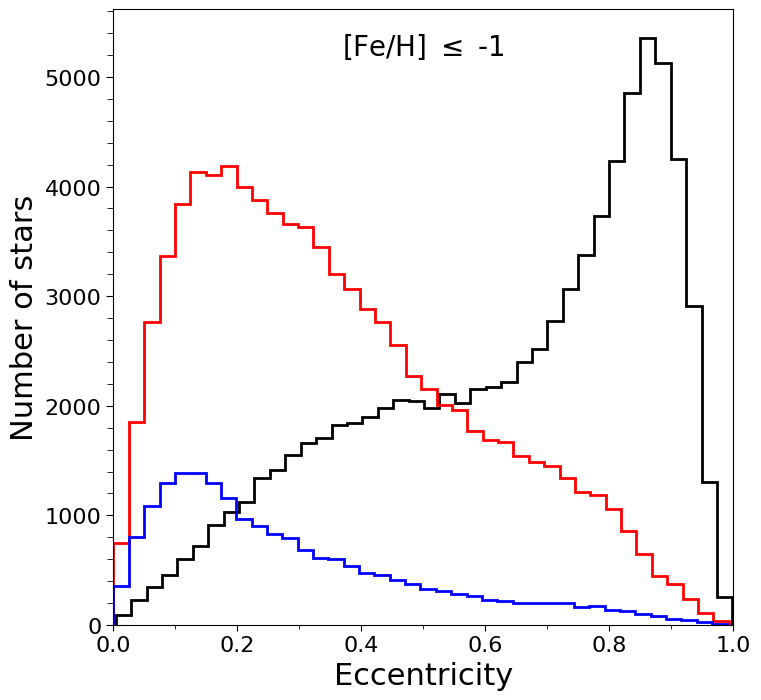}
    \caption{Left panel: cumulative number distribution of the MP ([Fe/H] $\leq -1$) stars as a function of [Fe/H], for stars with \zmax\ $> 3$ kpc (black line), \zmax\ $\leq 3$ kpc (red line), and \zmax\ $ \leq 1$ kpc (blue line). Middle panel: cumulative distribution functions of [Fe/H] for these subsamples. Each population is normalized on the basis of the number of stars at [Fe/H] $= -1$. Right panel: eccentricity distribution for these subsamples with [Fe/H] $\leq -1$.} 
    \label{Fig:vmpnumberfractionandecc}
\end{figure*}

\begin{figure*}
    \centering
    \includegraphics[width=\textwidth]{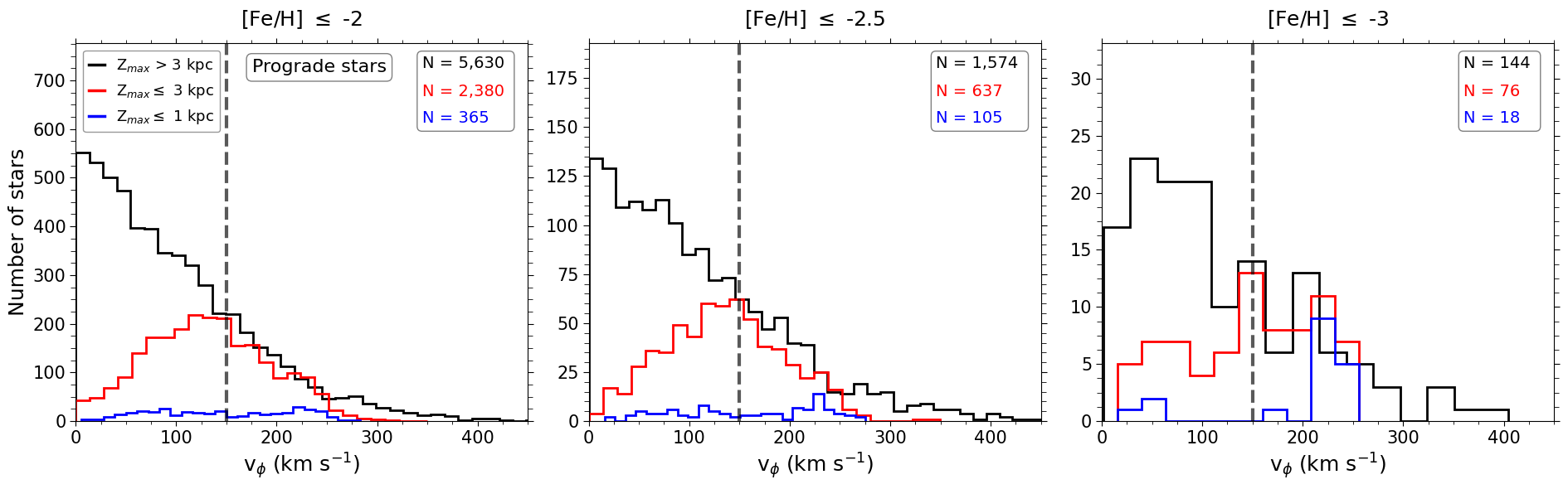}
    \caption{Number distributions of VMP/EMP stars with prograde orbits as a function of rotational velocity (v$_{\phi}$), for the stars with [Fe/H] $\leq -2$ (left panel), $\leq -2.5$ (middle panel), and $\leq -3$ (right panel), respectively. The vertical dashed line is at v$_{\phi} =$ 150 \kms, which is used to select disk-like stars on highly prograde orbits. Note that the bins on v$_{\phi}$ used for the stars with [Fe/H] $\leq -3$ are twice the size for the more metal-rich stars, due to their lower numbers.}
\label{Fig:vphihisto} 
\end{figure*}

\begin{figure*}
    \centering
    \includegraphics[width=0.95\textwidth]{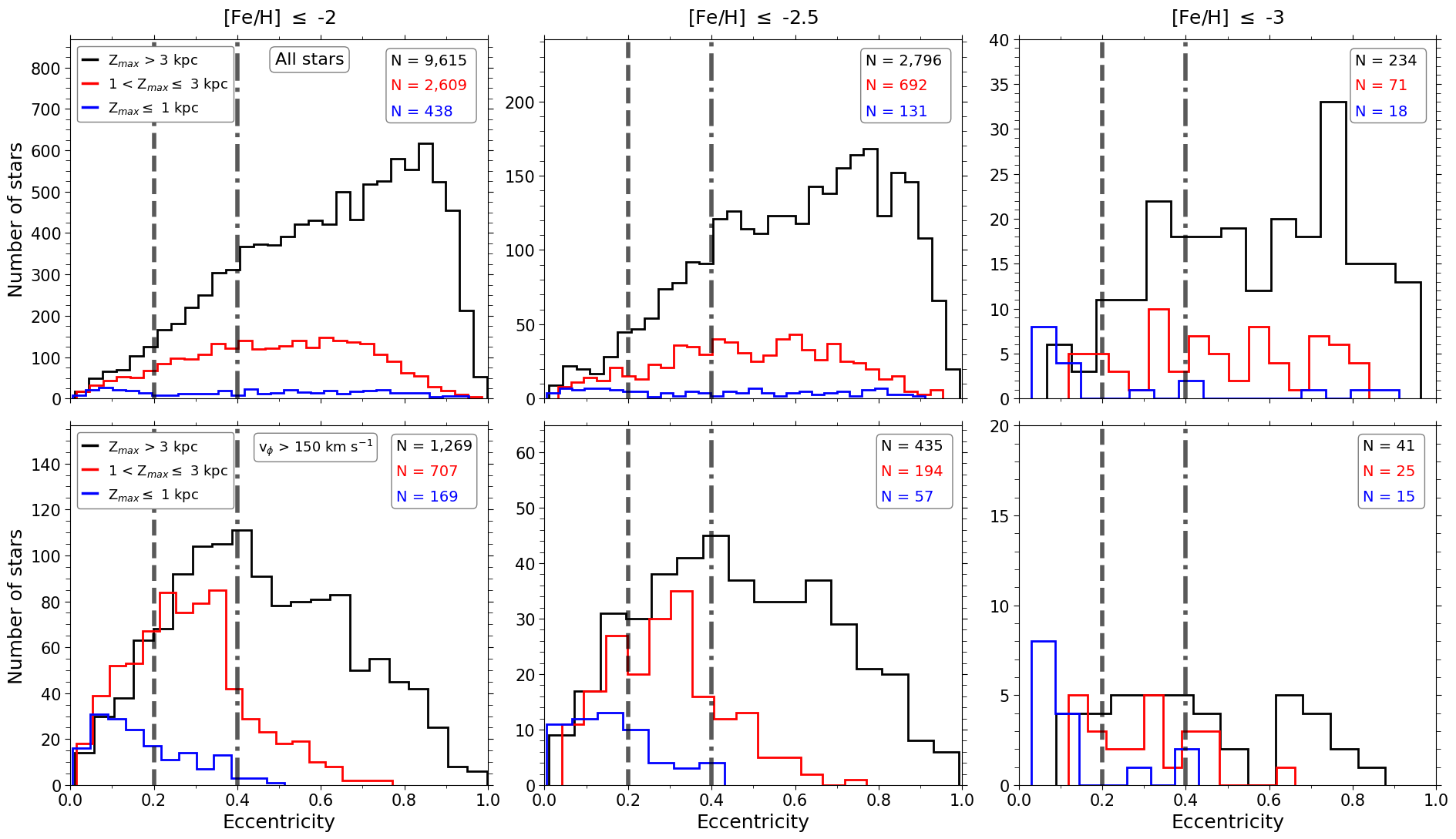}
    \caption{Top panels: number distributions of VMP/EMP stars on both retrograde and prograde orbits as a function of eccentricity, from the left to right panel, for [Fe/H] $\leq -2$, $\leq -2.5$, and $\leq -3$, respectively. The black, red, and blue solid lines indicate the stars with \zmax\ $> 3$ kpc, 1 kpc $<$ \zmax\ $\leq 3$ kpc, and \zmax\ $ \leq 1$ kpc, respectively. The number of stars in each region is indicated in the legend in the upper right corner of the panels. The dashed and dotted-dashed lines are shown at ecc = 0.2 and 0.4, respectively. Bottom panels: same as the top panels, but for the highly prograde stars with v$_{\phi}$ $>$ 150 \kms.}
    \label{Fig:ecczmaxhisto}
\end{figure*}

\subsection{Fractions of Disk-like and Halo-like Stars}\label{sec:diskhalolikeresults}

\subsubsection{Based on the \zmax\ Criterion}\label{sec:zmaxcriterion}

We first consider MP (MP; [Fe/H] $\leq -1$) stars in three regions of \zmax: \zmax\ $>$ 3 kpc, \zmax\ $\leq$ 3 kpc, and \zmax\ $\leq$ 1 kpc . We assign the stars with \zmax\ $>$ 3 kpc to the halo populations,
while those with \zmax\ $\leq$ 3 kpc and \zmax\ $\leq$ 1 kpc are candidate members of the MP thick- and thin-disk systems, respectively. The left panel of Figure~\ref{Fig:vmpnumberfractionandecc} shows the cumulative numbers of each population. At [Fe/H] $\leq -2$, approximately 9,600 halo stars with \zmax\ $>$ 3 kpc (black line) were found, along with about 3,000 stars with \zmax\ $\leq$ 3 kpc (red line) and 430 stars with \zmax\ $\leq$ 1 kpc (blue line). The middle panel shows the cumulative distribution function (CDF) for each population, normalized by the number of MP stars. Roughly 11.5\% of the MP stars assigned to the halo system are VMP stars, and 0.24\% are EMP stars; about 3.5\% of the MP stars assigned to the disk system are VMP stars, and 0.08\% are EMP stars. It is interesting to note that the CDFs of the stars with \zmax\ $\leq$ 3 kpc and \zmax\ $\leq$ 1 kpc are almost identical.

The right panel of Figure~\ref{Fig:vmpnumberfractionandecc} shows the distribution of orbital eccentricity for [Fe/H] $\leq -1$ of these subsamples split on \zmax. The broad distribution of eccentricity, peaking at high eccentricity, for stars kinematically assigned to the halo population is clear, as is the presence of low-eccentricity stars among the stars assigned to the disk system. Note that at this point we have not applied any cuts on v$_{\phi}$, only on \zmax, so we expect that the subsamples of stars with \zmax\ $\leq$ 3 kpc and \zmax\ $\leq$ 1 kpc have some level of contamination from halo stars. 

Figure~\ref{Fig:vphihisto} shows the distribution of v$_{\phi}$ for these three subsamples, but only for stars with [Fe/H] $\leq -2$, [Fe/H] $\leq -2.5$, and [Fe/H] $\leq -3$, from the left to right panels, respectively. The vertical dashed line corresponds to a cut on v$_{\phi} = 150$ \kms, the average orbital rotation value for the MWTD from \citet{Carollo2010}. We note that the adopted limit for the MWTD stars with the lowest v$_{\phi}$ from Carollo et al. (as well as from \citealt{An2021b}) is v$_{\phi} \sim 100$ \kms. From inspection, there remains considerable contamination of the VMP/EMP stars with prograde disk-like orbits by stars with prograde halo-like orbits, even with the higher cut at v$_{\phi} > 150$ \kms\ (although it is substantially less for the EMP stars shown in the right panel), indicating that a more sophisticated separation methodology is desirable. 

Figure~\ref{Fig:ecczmaxhisto} shows histograms of the eccentricity distribution for stars with [Fe/H] $\leq -2$, $\leq -2.5$, and $\leq -3$, respectively. We now subdivide the stars into three regions: \zmax $> 3$ kpc, 1 kpc $<$ \zmax\ $\leq 3$ kpc, and \zmax\ $\leq 1$ kpc, in an attempt to better isolate stars with thick-disk orbits from those with thin-disk orbits.  We note that these divisions are imperfect, in that we expect there to be contamination from halo stars at all \zmax. Within the 1 kpc $<$ \zmax\ $\leq 3$ kpc region there should be few 
thin-disk stars. Within the cut \zmax\ $ \leq 1$ kpc there will also remain some contamination from thick-disk stars.

From inspection of the top row of panels in Figure~\ref{Fig:ecczmaxhisto}, which includes stars on both retrograde and prograde orbits, the VMP/EMP stars do not exhibit prominent low-eccentricity orbits in all three ranges of [Fe/H]. However, in the bottom row of panels,
the introduction of the v$_{\phi} > 150$ \kms\ cut greatly increases the relative dominance of VMP/EMP stars with disk-like orbits, including for stars with ecc $\leq 0.4$. We note that similar results for VMP stars have been found by \citet{Bellazzini2024}, based on a sample of some 700,000 stars with photometric-metallicity estimates obtained with synthetic Str\"omgren photometry from $Gaia$ DR3 by \citet{Bellazzini2023}.

\subsubsection{Based on the Haywood Criterion}\label{sec:haywoodcriterion}

\begin{figure*}
    \centering
    \includegraphics[width=\textwidth]{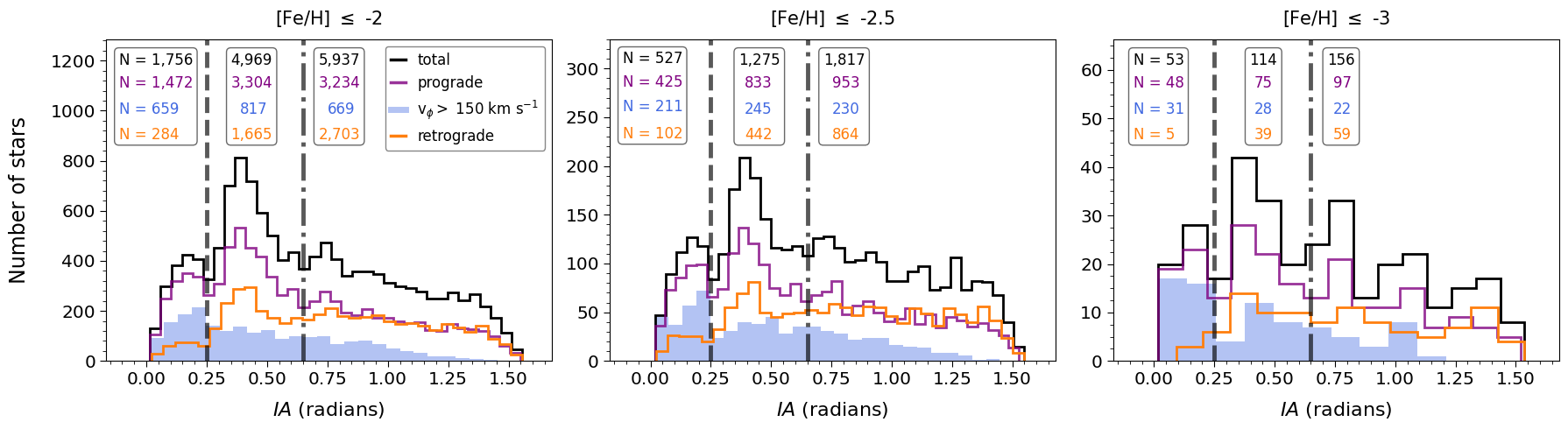}
    \caption{Number distribution of VMP/EMP stars as a function of $IA$, for [Fe/H] $\leq -2$, $\leq -2.5$, and $\leq -3$, from the left to right, respectively. The total, prograde, and retrograde orbiting stars are shown with black, purple, and orange lines, respectively. The prograde stars with v$_{\phi} >$ 150 \kms\ are shaded in blue. The number of stars in each region is indicated in the legend on the top of the panels.  The dashed and dotted-dashed lines indicate $IA$ = 0.25 and 0.65 rad, respectively.}
    \label{Fig:arcthisto}
\end{figure*}

There is also evidence for the existence of a VMP/EMP disk system from the Haywood 
diagram. Figure~\ref{Fig:arcthisto} shows the distribution of the arctangent of the \zmax/\Rmax\ values (defined here as the inclination angle, $IA$) for VMP/EMP stars, following \citet{Haywood2018} and \cite {DiMatteo2020}. The panels show this distribution for stars with the total numbers of VMP/EMP stars (black line), for stars with prograde orbits (purple line), and for stars with retrograde orbits (orange line), respectively, for stars in the regions with [Fe/H] $\leq -2$, $\leq -2.5$, and $\leq -3$. The vertical dashed line and dotted-dashed line show the approximate ``troughs" in the distributions at $IA$ = 0.25 and 0.65 rad, respectively, which can be used to roughly separate likely halo stars, thick-disk stars, and thin-disk stars. The numbers shown in each region listed in the figure reveal that VMP/EMP stars with prograde orbits dominate over those with retrograde orbits for $IA \leq 0.25$ rad and those with $0.25$ rad $< IA \leq 0.65$ rad, and much less so for $IA > 0.65$ rad. One can reasonably associate the prograde stars with $IA \leq 0.25$ rad with thin-disk orbits, those with $0.25$ rad $< IA \leq 0.65$ rad with thick-disk orbits, and those with $IA > 0.65$ rad with halo-like orbits.

If we now specialize to the highly prograde stars with orbital velocities v$_{\phi} > 150$ \kms (indicated by the blue shaded region in Figure~\ref{Fig:arcthisto}), the relative dominance of the stars in the disk-like system for VMP/EMP stars becomes even clearer. 

\begin{figure*}
    \centering
    \includegraphics[width=0.95\textwidth]{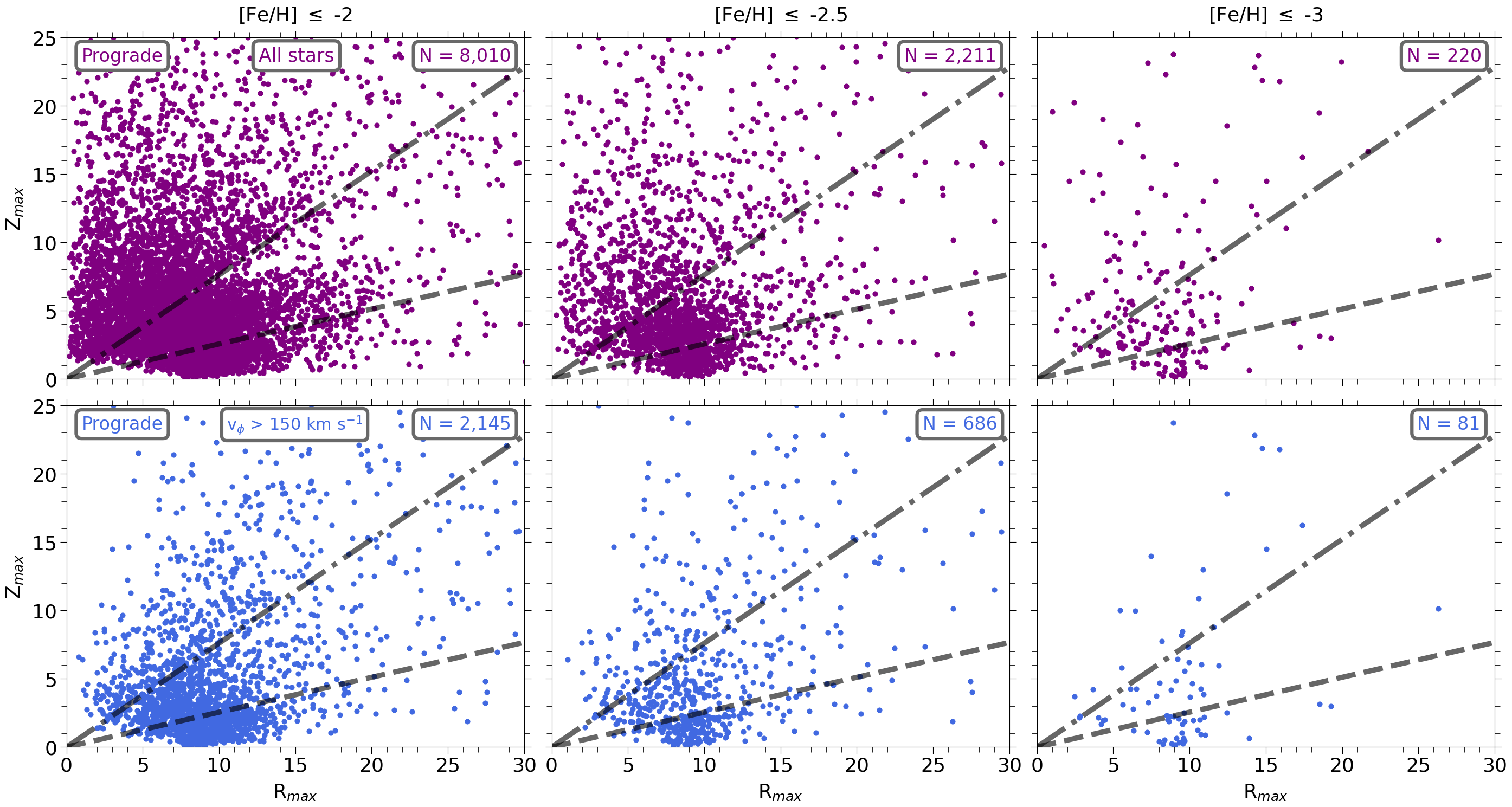}
    \caption{Top panels: the \zmax\ distribution as a function of \Rmax\ for the full sample of prograde stars. Dashed and dotted-dashed lines represent $IA$ = 0.25 and 0.65 rad, respectively. The number of stars is shown in the legend at the top right of each panel. Bottom panels: same as the top panels, but for the stars with v$_{\phi} >$ 150 \kms.}
    \label{Fig:rmaxzmaxdistrib}
\end{figure*}

Figure~\ref{Fig:rmaxzmaxdistrib} is a plot of \zmax\ vs. \Rmax\ for the stars with [Fe/H] $\leq -2$, $\leq -2.5$, and $\leq -3$, in the left, middle, and right panels, respectively. The top panels of this figure show plots of the \zmax\ distribution as a function of \Rmax\ for the full sample of prograde stars (v$_{\phi} >$ 0 \kms.) The dashed and dotted-dashed lines correspond to the troughs shown in Figure~\ref{Fig:arcthisto} at $IA$ = 0.25 and 0.65 rad, respectively. The number of stars is provided in the legend at the top of each panel. The bottom panels apply to the stars on highly prograde orbits (v$_{\phi} >$ 150 \kms).  

\begin{figure*}
    \centering
    \includegraphics[width=0.94\textwidth]{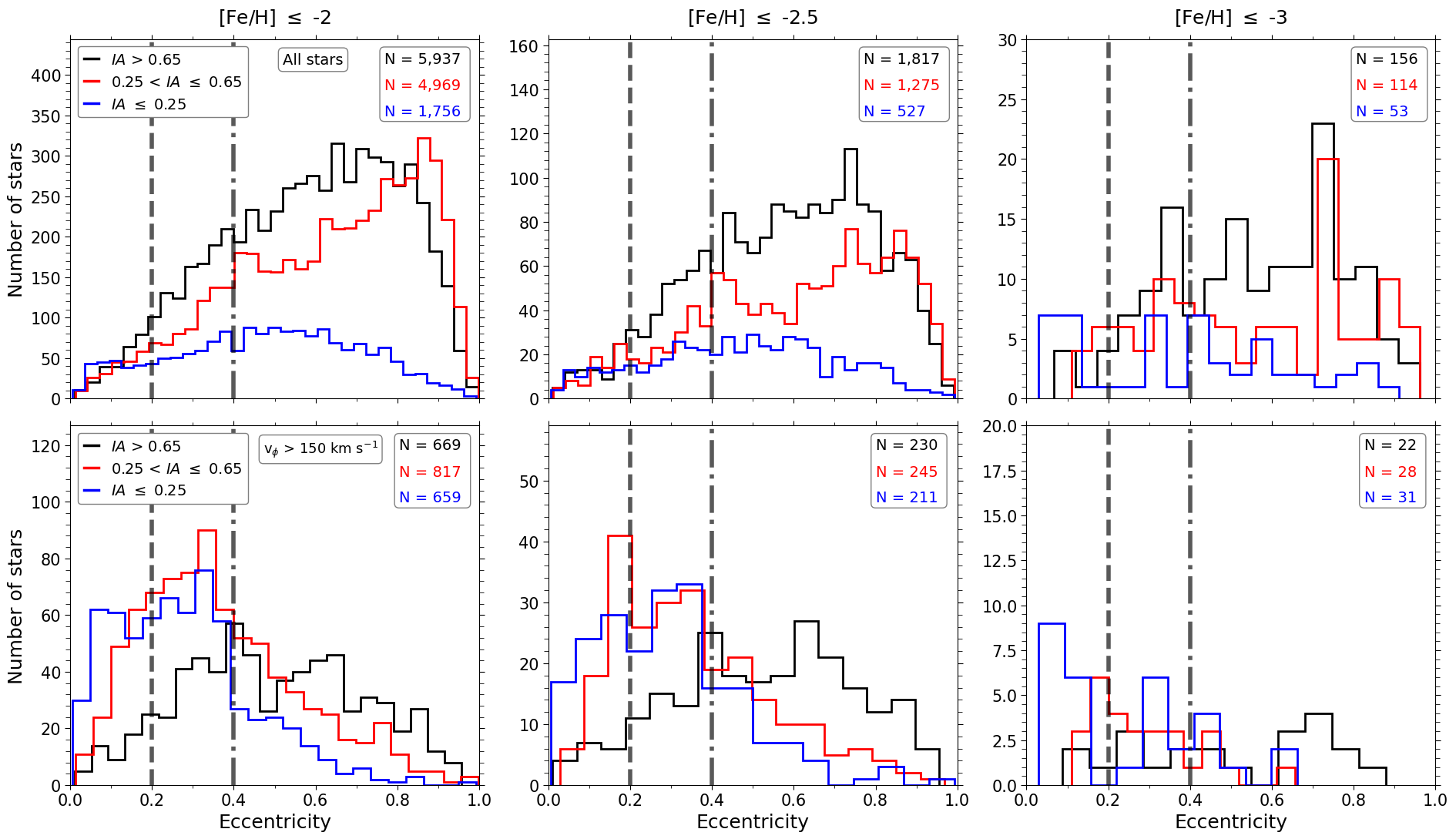}
    \caption{Top panels: Number distributions of VMP/EMP stars as a function of eccentricity, from the left to the right panel, for [Fe/H] $\leq −2$, $\leq −2.5$, and $\leq −3$, respectively. The black, red, and blue solid lines indicate the stars with $IA > 0.65 $, $0.25 < IA \leq 0.65$, and $IA \leq 0.25$ radians, respectively. The dashed and dot-dashed lines are shown at eccentricity = 0.2 and 0.4, respectively. The number of stars in each region is indicated in the legend at the top right of the panels. Bottom panels: Same as in the top panels, but for the highly prograde stars with v$_{\phi} > 150$ km s$^{-1}$.}
    \label{Fig:eccarcthisto}
\end{figure*}

Figure~\ref{Fig:eccarcthisto} shows histograms of the eccentricity distribution for stars with [Fe/H] $\leq -2$, $\leq -2.5$, and $\leq -3$, respectively. The colors represent the same cuts on $IA$ as in Figure~\ref{Fig:arcthisto}: halo-like orbits with $IA > 0.65$ rad are shown with black lines, thick-disk orbits with $0.25$ rad $< IA \leq 0.65$ rad are shown with red lines, and thin-disk orbits with $IA \leq 0.25$ rad are shown with blue lines. From inspection of the top row of panels, which includes stars on both retrograde and prograde orbits, the candidate VMP/EMP thin disk-like stars are broadly distributed over all eccentricities,  while the candidate VMP/EMP thick disk-like stars exhibit similar patterns to halo-like stars in all three ranges of [Fe/H]. However, in the bottom row of panels, the v$_{\phi} > 150$ \kms\ cut increases the relative dominance of VMP/EMP stars on disk-like orbits, including for stars with ecc $\leq 0.4$.

In summary, a total of 1,496 candidate VMP/EMP stars with v$_{\phi} >$ 150 \kms\ are identified.  The total numbers of highly prograde disk-like candidates selected by the \zmax$~$and Haywood criteria are 876 and 1,476, respectively. There are 856 stars selected by both methods. 

\vskip 1.5cm
\section{Discussion}\label{sec:discussion}

Separation based on the \zmax\ criterion, although capable of identifying a relatively pure sample of stars with halo-like orbits, has considerable potential contamination of stars with disk-like orbits by halo-like stars. The Haywood criterion, based on the separation of stars in different dynamical populations, produces a purer sample of disk-like stars, especially when the dominance of stars with thick and thin disk-like orbits over stars with halo-like orbits at ecc $< 0.4$ is considered in conjunction.

As we have demonstrated in this paper, the large numbers of stars now available with photometric-metallicity estimates from SMSS and SAGES have increased the numbers of candidate disk system VMP/EMP stars dramatically. 

We refer to these stars as candidates for two reasons. First, although the photometric-metallicity estimates have a precision of $\sim$ 0.1\,dex for [Fe/H] $\geq -1$ and approximately 0.3\,dex at [Fe/H] $\sim -3.5$ to $-4.0$, comparable to those obtained from low to medium resolution ($R=\lambda / \Delta \lambda \sim 1,800$) with a signal-to-noise ratio greater than 20 \citep{Luo2015,Yanny2009,Rockosi2022}, they may be influenced by the presence of strong molecular carbon bands, in particular for the most metal-poor stars. We have taken steps to mitigate this behavior, as described in the Appendix, but they should be confirmed by follow-up spectroscopy. Secondly, the question remains whether at least some of the apparent disk system VMP/EMP stars represent members of an early forming \textit{in-situ} primordial disk system, prior to additional stars being added from accreted dwarf satellites, or are possibly a very/extremely low metallicity tail of the long-recognized MWTD component of the MW. These alternatives may prove difficult to differentiate between based on kinematics alone, as mergers with dwarf galaxies could readily perturb the orbits of stars that were born in a primordial thin or thick disk. 

The best way to distinguish between these two possibilities may be to conduct a thorough study of their elemental abundances and look for differences as a function of declining metallicity. \citet{Feltzing2023} have recently used elemental abundance information from APOGEE, in combination with kinematics, in order to identify the likely presence of an early disk structure in the inner disk of the MW including VMP stars (although they are limited by the lack of lower-metallicity stars in APOGEE, precluding verification that EMP stars are present as well).  Detailed chemical abundances for our candidate VMP/EMP stars would clearly be useful.  

Additional information should soon be available from the J-PLUS and S-PLUS photometric surveys, which can obtain estimates for C and Mg (as well as N and Ca, once ongoing calibrations are completed), in addition to [Fe/H], thanks to their narrow/medium-bandpass filters.  More complete information will require high-resolution spectroscopic follow-up for at least a subset of the VMP/EMP candidates. Determination of more accurate age estimates than we have at present for candidate VMP/EMP stars on disk-like orbits may also prove illuminating. 

\begin{figure}
    \centering
    \includegraphics[width=0.45\textwidth]{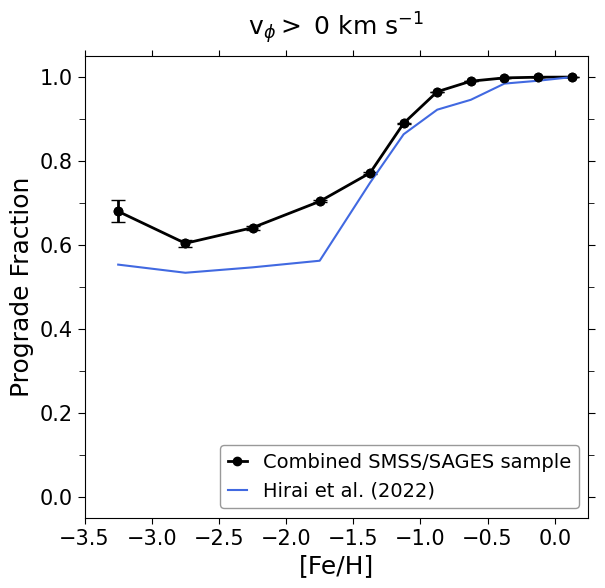}
    \caption{Fraction of prograde stars as a function of [Fe/H]. The black line shows the combined SMSS/SAGES sample. The blue line shows the simulation results of \citet{Hirai2022}. The Galactocentric distance of the data considered is confined to between 3 and 20 kpc, roughly corresponding to the combined SMSS/SAGES sample. The error bars shown are calculated using the normal approximation for the binomial proportions; they are quite small owing to the large number of stars in our data set.}
    \label{Fig:progradefraction}
\end{figure}

\subsection{Comparison with Simulations}\label{sec:Simulations}

Beyond the identification of the VMP/EMP disk system candidates, we can speculate on their origins by considering numerical simulations of an MW-like galaxies. We analyzed the data from a high-resolution cosmological zoom-in simulation of a Milky Way-like galaxy with a halo mass of $1.2\,\times\,10^{12}\ M_{\sun}$ presented in \citet{Hirai2022}. 
These authors defined the \textit{in-situ} component as stars formed in the main halo of the central galaxy, whereas the accreted component was defined as stars coming from dwarf galaxy satellites. Data within the Galactocentric distance $r_{\GC}$ between 3 and 20 kpc were considered; this region roughly corresponds to the observed region by SMSS and SAGES. 

From this simulation, we found that 8\% and 92\% of stars with v$_{\phi}>$ 150 \kms\ and [Fe/H] $\leq -2$ are formed in the \textit{in-situ} and accreted components, respectively. We also found that 96\% of VMP/EMP stars with v$_{\phi}>$ 150 \kms\ have ages $>$ 10 Gyr. 

Similar results have been shown in the analysis of \texttt{IllustrisTNG50} simulations by \citet{Mardini2022a} and \citet{Carollo2023}. Most recently, \citet{Sotillo-Ramos2023} considered a large sample of 138 MW analogs from the TNG50 cosmological simulations and found that, across all of these analogs, about 20\% of the VMP/EMP stars have disk-like orbits, with some analogs reaching as high as 30\%.  Roughly half of their disk-like stars have average ages exceeding 12.5 Gyr, with 70\% coming from accreted dwarf galaxies.
Taken as a whole, the simulation results suggest that VMP/EMP stars with disk-like orbits comprise stars coming primarily from accreted dwarf galaxies and \textit{in-situ} stars formed in an early primordial disk, or are associated with the MWTD.

Fractions of stars with prograde orbits can also inform the origin of VMP/EMP stars. Figure \ref{Fig:progradefraction} compares the prograde fractions as a function of [Fe/H] for the combined SMSS/SAGES sample and the simulation results of \citet{Hirai2022}. For [Fe/H] $>-2.0$, both our SMSS/SAGES sample and the simulation show an increasing trend toward higher metallicity, attributable to disk formation. On the other hand, the prograde fraction in our sample is roughly constant as a function of [Fe/H] for VMP/EMP stars. In the lowest-metallicity regime, the fraction rises to 0.68, reflecting the higher fraction of disk-like orbits among EMP stars (Figures \ref{Fig:ecczmaxhisto} and \ref{Fig:eccarcthisto}). This tendency is not clearly seen in the simulation. However, note that our sample's prograde fraction for VMP/EMP stars is significantly larger than that of the simulation.  It should be kept in mind that the \citet{Hirai2022} simulation is for a single realization, when in fact a variety of galaxy assembly histories are likely to have different outcomes, as demonstrated by other recent simulation studies \citep[e.g.,][]{Santistevan2021}.

Prograde orbit fractions higher than 0.5 for [Fe/H] $<-2$ suggest the accretion of satellites preferentially on prograde orbits or early disk formation at low metallicity. \citet{Carter2021} also reported a high prograde fraction, between 0.7 and 0.8, but with significantly larger error bars owing to the smaller sample they considered (see the top left panel of their Figure 3). They have also shown that the prograde fraction converges to 0.5 with a model assuming an isotropic distribution of orbits in the stellar halo \citep{Rybizki2018}. We confirm their results with a larger sample. Recently, \citet{Li2022a} have shown that 10 out of 12 MW stellar streams with an average [Fe/H] $\approx -2$ are on prograde orbits. These enhanced prograde fractions mean that the MW's VMP/EMP stars tend to be formed in accreted components with prograde orbits or in an ancient disk. As discussed above, spectroscopic follow-up of our candidate disk-like VMP/EMP stars may help improve estimates of the relative fractions associated with these differing origins. 

For convenience of future comparisons of our observations with those of others and with numerical simulations, Table \ref{tab:numbers} provides a summary of the numbers, fractions, and orbital characteristics of the SMSS/SAGES sample for different cuts on [Fe/H], \zmax, and $IA$.  Note that, except for the first line in each subsection of the table (indicated as ``All"), the fractions refer to the total numbers of stars listed on the first line at the top of each column in the subsection (shown in bold).

\include{numbers}

\section{Summary}\label{sec:summary}

We have identified 1,496 candidate VMP/EMP disk system stars in the MW from a subset of the $\sim 50$ million stars from SMSS and SAGES with available photometric-metallicity estimates, based on calibrated $u - G_{\rm BP}$ colors and $v - G_{\rm BP}$ colors, a combination of the $u/v$-bands from SMSS/SAGES and the ultra wide band $Gaia$ $G_{\rm BP}$ prism spectra (\citealt{Huang2022, Huang2023}).  We then trimmed the combined sample, eliminating photometrically identified binaries, cool dwarfs, and likely members of globular clusters. We then obtain the subset of 7.19 million stars in the combined sample with available RVs, proper motions, and distance estimates.

After the determination of dynamical parameters, we remove likely unbound stars and excise stars with errors in their orbital rotation velocities v$_{\phi} > 25$ \kms\, and relative errors in \zmax\ (maximum orbital distance from the Galactic plane) and in \Rmax\ (maximum orbital apocentric distance projected on to the plane) $> 30\%$, leaving a total sample of about 5.86 million stars. 

We then apply two methods to separate stars with halo-like and disk-like orbits. The first approach considered stars with \zmax $> 3$ kpc to have halo-like orbits and those with \zmax\ $\leq 3$ kpc to have disk-like orbits. Our analysis indicates that there exists a significant population of candidate VMP/EMP disk system stars, moving on rapid prograde orbits (v$_{\phi} > 150$ \kms), increasing their relative populations with declining metallicity. We also split the stars with disk-like orbits into the regions 1 $<$ \zmax\ $\leq 3$ kpc, and \zmax\ $\leq 1$ kpc, in an attempt to better isolate stars with thick-disk orbits from those with thin-disk orbits. Based on this criterion, we find that 28.7\% of the VMP stars with \zmax $\leq 3$ kpc have highly prograde disk-like orbits (707 stars on thick-disk orbits, 169 stars on thin-disk orbits), while 44.9\% of the EMP stars have highly prograde disk-like orbits (25 on thick-disk orbits, 15 on thin-disk orbits). These fractions increase further if one also takes the eccentricity of the orbits into account.

The second approach considered the stars populating wedges in the diagram of \zmax\ vs. 
$IA$, which redistributes corresponding to different dynamical populations of stars with halo-like and disk-like orbits. Our analysis indicates that there exists a significant population of candidate VMP/EMP disk system stars moving on rapid prograde orbits (v$_{\phi} > 150$ \kms), increasing their relative populations with declining metallicity. Based on the Haywood criterion, we find that 21.9\% of the VMP stars have highly prograde disk-like orbits (817 stars on thick-disk orbits, 659 stars on thin-disk orbits), while 35.3\% of the EMP stars have highly prograde disk-like orbits (28 stars on thick-disk orbits, 31 stars on thin-disk orbits). These fractions increase further if one also takes the eccentricity of the orbits into account. 

In the near future, the astrophysical properties and origin of these stars will be examined further with data from the large-scale Javalambre/Southern Photometric Local Universe Surveys (J/S-PLUS). These surveys include additional narrow/medium-band filters that allow for photometric estimates of C, N, Mg, and Ca abundances, once ongoing calibration efforts are completed. Of importance, it will then be possible to greatly reduce the influence of carbon on the metallicity estimates, which affect our current SMSS/SAGES sample, as [C/Fe] can be estimated separately from the [Fe/H].  The accuracy and precision of the derived metallicities will be improved as well.

In order to confirm the metallicities and elemental abundance estimates (such as the $\alpha$-elements or carbonicity, [C/Fe]) for the VMP/EMP stars with disk-like orbits, we require medium-resolution spectroscopic follow-up for the catalog of $\sim$ 1,500 VMP/EMP stars in our sample. High-resolution spectroscopic follow-up of at least the most interesting subset of these would also be useful. The full catalog of VMP/EMP stars with disk-like orbits is listed in the Appendix and will be made available in the online material. Determinations of age estimates for our candidate VMP/EMP stars would also help to place constraints on their origins.  Nevertheless, our present finding that large fractions of VMP/EMP stars are kinematically associated with the rapidly rotating MW disk system (in particular those at low eccentricity) strongly suggests the presence of an early forming ``primordial" disk.

\vspace{2.0cm}
\begin{acknowledgements}

The authors express thanks to Deokkeun An for his thoughts and comments on an early version of this paper.  We are also grateful to Evan Kirby and Borja Anguiano for their input as well. The authors thank an anonymous referee for their helpful comments that greatly improved this paper.

The Stellar Abundance and Galactic Evolution Survey (SAGES) is a multiband photometric project built and managed by the Research Group of the Stellar Abundance and Galactic Evolution of the National Astronomical Observatories, Chinese Academy of Sciences (NAOC). 

This work was supported in part by grant PHY 14-30152, Physics Frontier Center/JINA Center for the Evolution of the Elements (JINA-CEE), and by OISE-1927130: The International Research  Network for Nuclear Astrophysics (IReNA), awarded by the US National Science Foundation. 
Y.S.L. acknowledges support from the National Research Foundation (NRF) of Korea grant funded by the Ministry of Science and ICT (NRF-2021R1A2C1008679 and RS-2024-00333766). Y.S.L. also gratefully
acknowledges partial support for his visit to the University of Notre Dame from OISE-1927130: The International Research Network for Nuclear Astrophysics (IReNA), awarded
by the US National Science Foundation. 
Y. Huang acknowledges support from the National Key R\&D Program of China No. 2019YFA0405500 and National Natural Science Foundation of China grants 11903027 and 11833006.
Y. Hirai was supported by JSPS KAKENHI Grant Numbers JP22KJ0157, JP21H04499, JP21K03614, and JP22H01259. Numerical computations were in part carried out on Cray XC50 and computers at the Center for Computational Astrophysics, National Astronomical Observatory of Japan.
S.X. acknowledges support from the National Natural Science Foundation of China through project NSFC 12222301.
K.T. was supported by the National Natural Science Foundation of China under grant Nos. 12261141689, 12090044 and 12090040.
G.Z. acknowledges support by the National Natural Science Foundation of China under grant nos. 11988101 and 11890694.

\end{acknowledgements}

\vfill\eject

\gdef\thefigure{\thesection.\arabic{figure}}    

\bibliography{main}{}
\bibliographystyle{aasjournal}

\begin{appendix}

\setcounter{figure}{0}
\setcounter{table}{0}
\renewcommand{\thetable}{A\arabic{table}}
\renewcommand{\thefigure}{A\arabic{figure}}

Table~\ref{tab:colsfinalcatalog} provides a description of the parameters we report for candidate VMP/EMP stars from the combined SMSS/SAGES sample, based on the information provided by \citet{Huang2022, Huang2023}. We included stars with adopted photometric-metallicity estimates in the range $-4.0 < {\rm [Fe/H]} \leq -2.0$, based on the individual $u$-band and $v$-band filters, as well as their combination. Note that, for completeness, we have included information for all of the stars with available photometric-metallicity estimates, regardless of their metallicities obtained by any filters, errors in their derived metallicities, or reddening. The full table of 1,291,424 stars is made available in the online material.

Table~\ref{tab:finalcatalog} is a listing of the candidate VMP/EMP stars on disk-like orbits stars we employ, providing the data needed for further analysis and/or spectroscopic follow-up observations. 
From the candidates of Table~\ref{tab:colsfinalcatalog}, we only included the stars with 
errors in their adopted metallicities err$_{\rm [Fe/H]} \leq 0.5$\,dex and stars with a difference of less than $\pm 0.5$\,dex between the $u$-band- and $v$-band-based abundances ($|\rm[Fe/H]_{ub} - \rm[Fe/H]_{vb}|$). For the reddening cut, the numbers of stars that would be excluded, depending on $E(B-V) \le$ 0.1, 0.2, 0.3, 0.4, and 0.5, are approximately 1.72 million, 0.23 million, 17,000, 2, and 0, respectively. The numbers of VMP/EMP stars that would be removed by these cuts are about 3,000, 500, 5, and no VMP/EMP stars for $E(B-V) >$ 0.4. Thus, we chose to only include the stars with $E(B-V) \le 0.3$ for our analysis. We also included stars having derived errors in their orbital rotation velocities v$_{\phi}$ $\leq 25$ \kms, and relative errors in \zmax\ $\leq 0.30$ and \Rmax\ $\leq 0.30$.

\begin{figure*}
    \centering
    \includegraphics[width=0.69\textwidth]{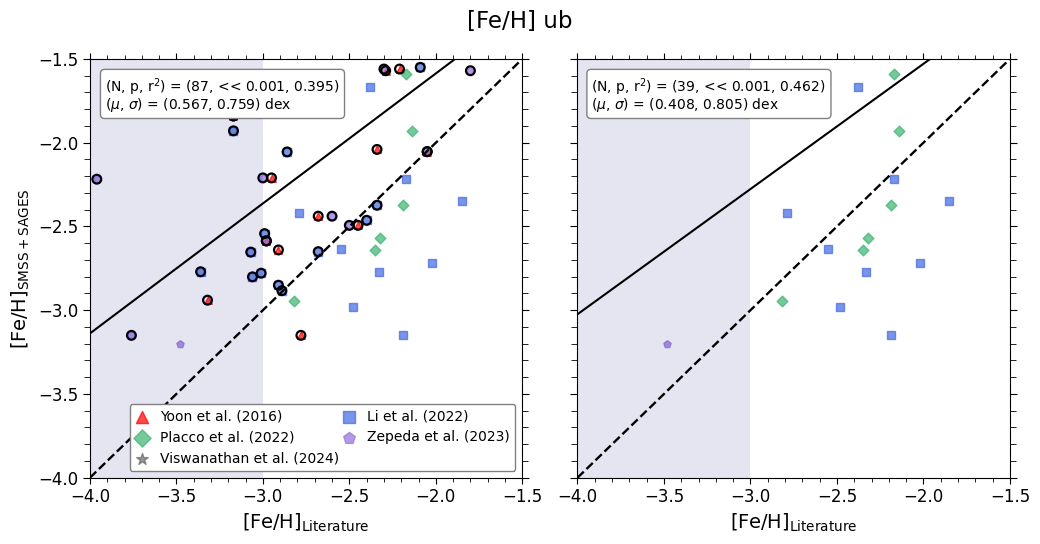}
    \includegraphics[width=0.69\textwidth]{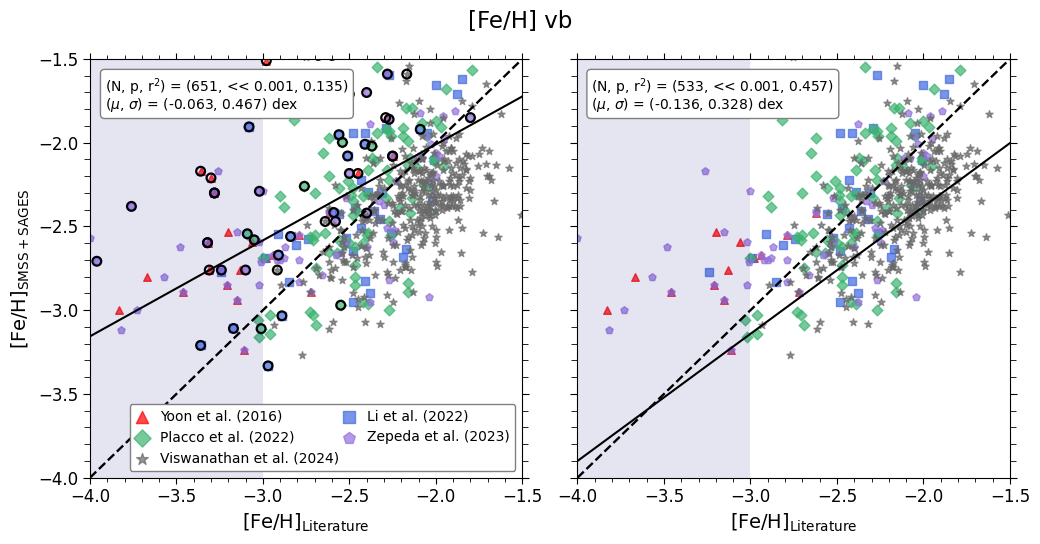}
    \includegraphics[width=0.69\textwidth]{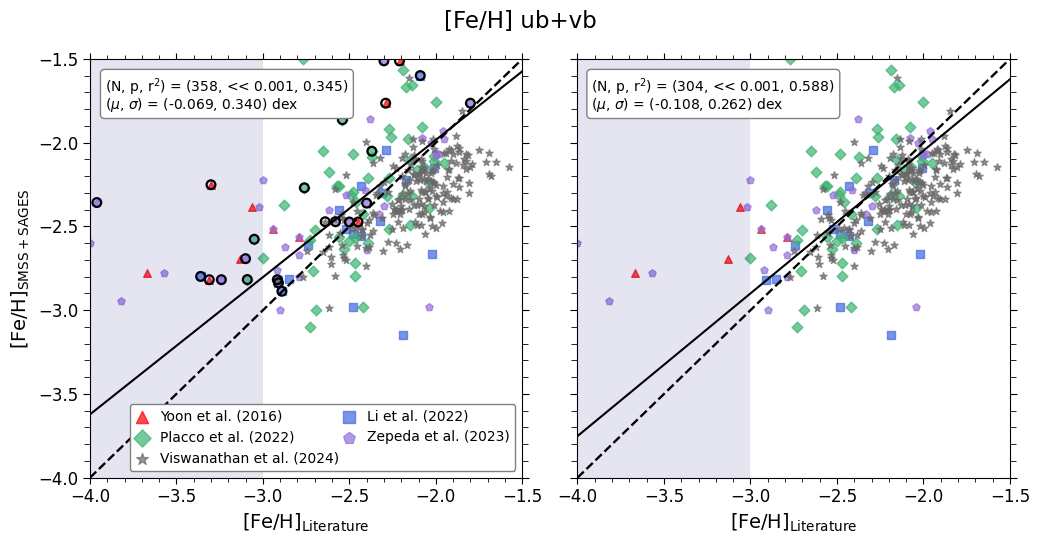}  
     \caption{Left column: comparison of the combined SMSS/SAGES photometric-metallicity estimates and spectroscopic metallicities for VMP/EMP stars, based on cross-matches to medium/high-resolution spectroscopic samples with available [Fe/H] and [C/Fe].  The stars from \citet{Yoon2016}, \citet{Li2022b}, \citet{Placco2022}, \citet{zepeda2023}, and \citet{Viswanathan2024} are shown as red triangles, blue squares, green diamonds, purple pentagons, and gray stars, respectively. The black circles indicate carbon-enhanced metal-poor (CEMP) stars that satisfy [C/Fe] $> +0.7$. Note that for our present purpose we employ the ``as observed" [C/Fe], without applying evolutionary corrections (e.g., from \citealt{Placco2014}). The black solid line is a linear regression line for all metallicity regions except for the range [Fe/H]$_{\rm Literature} \leq-3$, a light-blue shaded region, and the dashed line represents a one-to-one line. The legends in each panel indicate the number of matching stars (N), the $p$-value and the $r^{2}$ value in the Pearson correlation analysis, as well as the biweight location ($\mu$) and scale ($\sigma$) of the metallicity residuals. 
     Right column: comparison with the same spectroscopic catalogs, but excluding (recognized) CEMP stars. 
     From top to bottom, the panels indicate matches for stars with metallicities based on the  
    calibrated $u - G_{\rm BP}$ colors, $v - G_{\rm BP}$ colors, and when available, an average of both of these colors \citep{Huang2022, Huang2023}.}
    \label{Fig:ubvb_spec}
\end{figure*}

Figure \ref{Fig:ubvb_spec} compares our photometric-metallicity estimates (based on the $u - G_{\rm BP}$ colors, $v - G_{\rm BP}$ colors, and when available, the combination of these colors; see \citealt{Huang2022, Huang2023}) to medium- and high-resolution spectroscopic estimates with available [Fe/H] and [C/Fe] (not corrected for evolutionary effects) from a number of literature sources, including bright stars from $Gaia$ DR3 with spectroscopic metallicity estimates obtained by \citet{Viswanathan2024}, based on a refined analysis of the Radial Velocity Spectrometer (RVS) spectra.  The left column of panels shows the results from the full set of available stars in our catalog, while the right column of panels excludes the (recognized) carbon-enhanced metal-poor \citep[CEMP,][]{Beers2005} stars that satisfy [C/Fe] $> +0.7$. The black solid line in each panel is a linear regression for the metallicity region, excluding stars with [Fe/H]$_{\rm Literature}
\leq −3.0$, indicated with the light-blue shaded region. The dashed lines represent the one-to-one lines. The legends in each panel indicate the number of matching stars (N), the $p$-value, and the r$^{2}$ value (which indicates the fraction of variance that can be accounted for by the regression relationship) found by the Pearson correlation analysis, as well as the biweight location ($\mu$) and scale ($\sigma$) of the metallicity residuals \citep[see][]{Beers1990a}. The top, middle, and bottom panels apply to matching stars with available $\rm[Fe/H]_{ub}, \rm[Fe/H]_{vb}$ and $\rm[Fe/H]_{ub+vb}$, respectively.

From inspection of this figure, it is apparent that excluding (recognized) CEMP stars results in linear regression lines that are aligned more closely to the one-to-one relationships, particularly in the middle and bottom panels\footnote{Note that many of the literature stars with which we compare do not have published estimates of [C/Fe], so there no doubt exist more CEMP stars in our sample than shown in the figure, in particular among the EMP stars.}. This trend is supported by the higher r$^{2}$ values and smaller biweight residual location offsets and scale values shown in the right column of panels. It is evident that stars with enhanced carbon result in higher derived photometric-metallicity estimates in our analysis. The largest deviations are found when considering the photometric-metallicity estimates based solely on the $u$ band (top row of panels). Smaller deviations are found when considering the photometric-metallicity estimates based solely on the $v$ band (middle row of panels). The combination of the $u$-band and $v$-band photometric-metallicity estimates, as seen from the bottom row of panels, somewhat mitigates the effects of carbon enhancement, resulting in acceptably small offsets and lower dispersions. However, it is clear that our photometric-metallicity estimates for stars with literature estimates of [Fe/H] are most likely to be higher when carbon is enhanced, in particular for EMP stars.

The above results motivate our choice to only include stars with photometric-metallicity estimates based on the stars for which acceptable estimates are obtained based either solely on the $v$ band or on the combination of the $u$ band and $v$ band, but excluding stars that have estimates based solely on the $u$-band.  From this comparison, and under the assumption that the spectroscopic estimates of [Fe/H] from multiple sources themselves can account for a ``sample to sample" scatter (arising from different assumptions made by the individual analyses) on the order of 0.15-0.20\,dex, the external errors of the photometric-metallicity estimates range from 0.20-0.35\,dex (and on the order of 0.10 to 0.15\,dex for stars more metal-rich than considered here).  Note that this is also driven, at least in part, by the scatter induced by the presence of carbon.

\include{colsfinalcatalog}


\include{finalcatalog_stub}


\end{appendix}

\end{document}

%% file: numbers.tex
\startlongtable
\renewcommand{\arraystretch}{0.99}
\LTcapwidth=\textwidth
\tabletypesize{\scriptsize}
\begin{longtable*}{lcccc}
\caption*{Numbers, Fractions, and Orbital Characteristics of MP/VMP/EMP Stars in the SMSS/SAGES Sample \label{tab:numbers}} \\
\hline \hline
\multicolumn{5}{c}{Full Sample of MP Stars\rule[-2.6ex]{0pt}{6.3ex}}\\
\hline
 & $\rm{[Fe/H]} \leq -$1 & $\rm{[Fe/H]} \leq -$2 & $\rm{[Fe/H]} \leq -$2.5 & $\rm{[Fe/H]} \leq -$3 \rule[-2.5ex]{0pt}{6.4ex} \\
\textbf{All ($N_{\rm tot} = 171,005$)} & \textbf{171,005 (100.0 \%)}  &  \textbf{12,662 (7.4 \%)}  &  \textbf{3,619 (2.1 \%)}  &  \textbf{323 (0.2 \%)}  \rule[-2ex]{0pt}{5ex}\\
Prograde & 136,364 (79.7 \%)  &  8,010 (63.3 \%)  &  2,211 (61.1 \%)  &  220 (68.1 \%) \\
Retrograde & 34,641 (20.3 \%)  &  4,652 (36.7 \%)  &  1,408 (38.9 \%)  &  103 (31.9 \%) \\
Dwarf & 50,082 (29.3 \%)  &  2,806 (22.2 \%)  &  804 (22.2 \%)  &  71 (22.0 \%) \\
Giant & 120,923 (70.7 \%)  &  9,856 (77.8 \%)  &  2,815 (77.8 \%)  &  252 (78.0 \%) \\
  & $-1 \geq \rm{[Fe/H]} > -2$ & $-2 \geq \rm{[Fe/H]} > -2.5$ & $-2.5 \geq \rm{[Fe/H]} > -3$ & $\rm{[Fe/H]} \leq -$3 \rule[-2.6ex]{0pt}{7.5ex}\\
\textbf{All ($N_{\rm tot} = 171,005$)} & \textbf{158,343 (92.6 \%)}  &  \textbf{9,043 (5.3 \%)}  &  \textbf{3,296 (1.9 \%)}  &  \textbf{323 (0.2 \%)}  \rule[-2ex]{0pt}{5ex}\\
Prograde & 128,354 (81.1 \%)  &  5,799 (64.1 \%)  &  1,991 (60.4 \%)  &  220 (68.1 \%) \\
Retrograde & 29,989 (18.9 \%)  &  3,244 (35.9 \%)  &  1,305 (39.6 \%)  &  103 (31.9 \%) \\
Dwarf & 47,276 (29.9 \%)  &  2,002 (22.1 \%)  &  733 (22.2 \%)  &  71 (22.0 \%) \\
Giant & 111,067 (70.1 \%)  &  7,041 (77.9 \%)  &  2,563 (77.8 \%)  &  252 (78.0 \%)  \rule[-4ex]{0pt}{0ex}\\
\hline
\multicolumn{5}{c}{Z$_{\rm max}$ Criterion Separation of Orbits for VMP/EMP Stars \rule[-2.7ex]{0pt}{6.8ex}}\\
\hline
$\rm{[Fe/H]} \leq -$2 & Z$_{\rm max} >$ 3 kpc & Z$_{\rm max} \leq$ 3 kpc & 1 kpc $<$ Z$_{\rm max} \leq$ 3 kpc & Z$_{\rm max} \leq$ 1 kpc \rule[-2.7ex]{0pt}{7.5ex}\\
\textbf{All ($N_{\rm tot} = 12,662$)} & \textbf{9,615 (75.9 \%)}  &  \textbf{3,047 (24.1 \%)}  &  \textbf{2,609 (20.6 \%)}  &  \textbf{438 (3.5 \%)}  \rule[-2ex]{0pt}{5ex}\\
Prograde & 5,630 (58.6 \%)  &  2,380 (78.1 \%)  &  2,015 (77.2 \%)  &  365 (83.3 \%) \\
Retrograde & 3,985 (41.4 \%)  &  667 (21.9 \%)  &  594 (22.8 \%)  &  73 (16.7 \%) \\
Highly prograde & 1,269 (13.2 \%)  &  876 (28.7 \%)  &  707 (27.1 \%)  &  169 (38.6 \%) \\
Highly prograde, ecc $\leq$ 0.4 & 552 (5.7 \%)  &  745 (24.5 \%)  &  582 (22.3 \%)  &  163 (37.2 \%) \\
Highly prograde, ecc $\leq$ 0.2 & 147 (4.8 \%)  &  316 (10.4 \%)  &  205 (7.9 \%)  &  111 (25.3 \%) \\
$\rm{[Fe/H]} \leq -$2.5 & Z$_{\rm max} >$ 3 kpc & Z$_{\rm max} \leq$ 3 kpc & 1 kpc $<$ Z$_{\rm max} \leq$ 3 kpc & Z$_{\rm max} \leq$ 1 kpc \rule[-3ex]{0pt}{8ex}\\
\textbf{All ($N_{\rm tot} = 3,619$)} & \textbf{2,796 (77.3 \%)}  &  \textbf{823 (22.7 \%)}  &  \textbf{692 (19.1 \%)}  &  \textbf{131 (3.6 \%)}  \rule[-2ex]{0pt}{5ex}\\
Prograde & 1,574 (56.3 \%)  &  637 (77.4 \%)  &  532 (76.9 \%)  &  105 (80.2 \%) \\
Retrograde & 1,222 (43.7 \%)  &  186 (22.6 \%)  &  160 (23.1 \%)  &  26 (19.8 \%) \\
Highly prograde & 435 (15.6 \%)  &  251 (30.5 \%)  &  194 (28.0 \%)  &  57 (43.5 \%) \\
Highly prograde, ecc $\leq$ 0.4 & 182 (6.5 \%)  &  209 (25.4 \%)  &  154 (22.3 \%)  &  55 (42.0 \%) \\
Highly prograde, ecc $\leq$ 0.2 & 59 (2.1 \%)  &  94 (11.4 \%)  &  56 (8.1 \%)  &  38 (29.0 \%) \\
$\rm{[Fe/H]} \leq -$3 & Z$_{\rm max} >$ 3 kpc & Z$_{\rm max} \leq$ 3 kpc & 1 kpc $<$ Z$_{\rm max} \leq$ 3 kpc & Z$_{\rm max} \leq$ 1 kpc \rule[-3ex]{0pt}{8ex}\\
\textbf{All ($N_{\rm tot} = 323$)} & \textbf{234 (72.4 \%)}  &  \textbf{89 (27.6 \%)}  &  \textbf{71 (22.0 \%)}  &  \textbf{18 (5.6 \%)}  \rule[-2ex]{0pt}{5ex}\\
Prograde & 144 (61.5 \%)  &  76 (85.4 \%)  &  58 (81.7 \%)  &  18 (100.0 \%) \\
Retrograde & 90 (38.5 \%)  &  13 (14.6 \%)  &  13 (18.3 \%)  &  0 (0.0 \%) \\
Highly prograde & 41 (17.5 \%)  &  40 (44.9 \%)  &  25 (35.2 \%)  &  15 (83.3 \%) \\
Highly prograde, ecc $\leq$ 0.4 & 21 (9.0 \%)  &  31 (34.8 \%)  &  18 (25.4 \%)  &  13 (72.2 \%) \\
Highly prograde, ecc $\leq$ 0.2 & 7 (3.0 \%)  &  19 (21.3 \%)  &  7 (9.9 \%)  &  12 (66.7 \%)  \rule[-4ex]{0pt}{0ex}\\
\hline
\multicolumn{5}{c}{Haywood Criterion Separation of Orbits for VMP/EMP Stars \rule[-2.8ex]{0pt}{7ex}}\\
\hline
$\rm{[Fe/H]} \leq -$2 & $\it{IA}$ $>$ 0.65 & $\it{IA}$ $\leq$ 0.65 & 0.25 $<$ $\it{IA}$ $\leq$ 0.65 & $\it{IA}$ $\leq$ 0.25 \rule[-3ex]{0pt}{8ex}\\
\textbf{All ($N_{\rm tot} = 12,662$)} & \textbf{5,937 (46.9 \%)}  &  \textbf{6,725 (53.1 \%)}  &  \textbf{4,969 (39.2 \%)}  &  \textbf{1,756 (13.9 \%)}  \rule[-2ex]{0pt}{5ex}\\
Prograde & 3,234 (54.5 \%)  &  4,776 (71.0 \%)  &  3,304 (66.5 \%)  &  1,472 (83.8 \%) \\
Retrograde & 2,703 (45.5 \%)  &  1,949 (29.0 \%)  &  1,665 (33.5 \%)  &  284 (16.2 \%) \\
Highly prograde & 669 (11.3 \%)  &  1,476 (21.9 \%)  &  817 (16.4 \%)  &  659 (37.5 \%) \\
Highly prograde, ecc $\leq$ 0.4 & 249 (4.2 \%)  &  1,048 (15.6 \%)  &  734 (14.8 \%)  &  532 (30.3 \%) \\
Highly prograde, ecc $\leq$ 0.2 & 59 (1.0 \%)  &  404 (6.0 \%)  &  186 (3.7 \%)  &  237 (13.5 \%) \\
$\rm{[Fe/H]} \leq -$2.5 & $\it{IA}$ $>$ 0.65 & $\it{IA}$ $\leq$ 0.65 & 0.25 $<$ $\it{IA}$ $\leq$ 0.65 & $\it{IA}$ $\leq$ 0.25 \rule[-3ex]{0pt}{8ex}\\
\textbf{All ($N_{\rm tot} = 3,619$)} & \textbf{1,817 (50.2 \%)}  &  \textbf{1,802 (49.8 \%)}  &  \textbf{1,275 (35.2 \%)}  &  \textbf{527 (14.6 \%)}  \rule[-2ex]{0pt}{5ex}\\
Prograde & 953 (52.4 \%)  &  1,258 (69.8 \%)  &  833 (65.3 \%)  &  425 (80.6 \%) \\
Retrograde & 864 (47.6 \%)  &  544 (30.2 \%)  &  442 (34.7 \%)  &  102 (19.4 \%) \\
Highly prograde & 230 (12.7 \%)  &  456 (25.3 \%)  &  245 (19.2 \%)  &  211 (40.0 \%) \\
Highly prograde, ecc $\leq$ 0.4 & 69 (3.8 \%)  &  322 (17.9 \%)  &  211 (16.5 \%)  &  163 (30.9 \%) \\
Highly prograde, ecc $\leq$ 0.2 & 18 (1.0 \%)  &  135 (7.5 \%)  &  67 (5.3 \%)  &  72 (13.7 \%) \\
$\rm{[Fe/H]} \leq -$3 & $\it{IA}$ $>$ 0.65 & $\it{IA}$ $\leq$ 0.65 & 0.25 $<$ $\it{IA}$ $\leq$ 0.65 & $\it{IA}$ $\leq$ 0.25 \rule[-3ex]{0pt}{8ex}\\
\textbf{All ($N_{\rm tot} = 323$)} & \textbf{156 (48.3 \%)}  &  \textbf{167 (51.7 \%)}  &  \textbf{114 (35.3 \%)}  &  \textbf{53 (16.4 \%)}  \rule[-2ex]{0pt}{5ex}\\
Prograde & 97 (62.2 \%)  &  123 (73.7 \%)  &  75 (65.8 \%)  &  48 (90.6 \%) \\
Retrograde & 59 (37.8 \%)  &  44 (26.3 \%)  &  39 (34.2 \%)  &  5 (9.4 \%) \\
Highly prograde & 22 (14.1 \%)  &  59 (35.3 \%)  &  28 (24.6 \%)  &  31 (58.5 \%) \\
Highly prograde, ecc $\leq$ 0.4 & 8 (5.1 \%)  &  44 (26.3 \%)  &  29 (25.4 \%)  &  22 (41.5 \%) \\
Highly prograde, ecc $\leq$ 0.2 & 2 (1.3 \%)  &  24 (14.4 \%)  &  9 (7.9 \%)  &  15 (28.3 \%)  \rule[-4ex]{0pt}{0ex}\\
\bottomrule
\\ \\
\end{longtable*}

%% file: colsfinalcatalog.tex
\startlongtable\begin{deluxetable*}{l  l  l}
\renewcommand{\arraystretch}{1.05}
\tabletypesize{\scriptsize}
\tablecaption{Description of the Candidate VMP/EMP Stars in the Combined SMSS/SAGES Sample \label{tab:colsfinalcatalog}}
\tablehead{\colhead{Field} \hspace{1.5cm} & \colhead{Description} \hspace{11.1cm} & \colhead{Unit} \hspace{0.5cm} } 
\startdata
GaiaDR3 & The Gaia DR3 Source ID [source\_id] & $-$\\
R.A. & The Right Ascension from SMSS DR2 and SAGES DR1 (J2000) & hours : minutes : seconds\\
Decl. & The Declination from SMSS DR2 and SAGES DR1 (J2000) & degrees : minutes : seconds\\
GCmag & The calibration-corrected G magnitude by \citet{Huang2022, Huang2023} for the Gaia DR3 [G\_C] & $-$\\
e\_GCmag & The calibration-corrected G magnitude uncertainty by \citet{Huang2022, Huang2023} for the Gaia DR3 [err\_G\_C] & $-$\\
BR0 & The intrinsic colors of (G$_{\rm BP}$ − G$_{\rm RP}$)$_{0}$ by \citet{Huang2022, Huang2023} [br0] & $-$\\
e\_BR0 & The intrinsic colors uncertainty of (G$_{\rm BP}$ − G$_{\rm RP}$)$_{0}$ by \citet{Huang2022, Huang2023} [err\_br0] & $-$\\
E(B-V) & The E(B $−$ V) from the extinction map of \citet{Schlegel1998}, corrected by \citet{Huang2022, Huang2023} [ebv\_sfd] & $-$\\
FeH-UB/VB/UVB & The photometric-metallicity estimates from \citet{Huang2022, Huang2023} [feh].\\
$-$ & The ``ub", ``vb", and ``ub+vb" indicate the stellar color(s) used in estimating {$\rm [Fe/H]$} by \citet{Huang2022, Huang2023} & $-$\\
e\_FeH-UB/VB/UVB & The photometric-metallicity estimates uncertainty from \citet{Huang2022, Huang2023} [err\_feh] & dex\\
Teff & The effective temperature from \citet{Huang2022, Huang2023} [Teff] & K\\
e\_Teff & The effective temperature uncertainty from \citet{Huang2022, Huang2023} [err\_Teff] & K\\
Dist & The distance from \citet{Huang2022, Huang2023} [dist\_adop] & kpc\\
e\_Dist & The distance uncertainty from \citet{Huang2022, Huang2023} [err\_dist\_adop] & kpc\\
f\_Dist & Flag with ``parallax" if the distance is derived by parallax, \\
$-$ & ``CAF", ``CMD\_dwarf", ``CMD\_dwarf\_nobia", and ``CMD\_giant" if the distance is derived by \\
$-$ & color-absolute magnitude fiducial relations from \citet{Huang2022, Huang2023} [dist\_adop\_flg] & $-$\\
RVel & The radial velocity from \citet{Huang2022, Huang2023} [rv\_adop] & km s$^{-1}$\\
e\_RVel & The radial velocity uncertainty from \citet{Huang2022, Huang2023} [err\_rv\_adop] & km s$^{-1}$\\
f\_RVel & Flag with ``GaiaDR3", ``GALAH", ``LM-DR9", ``APG-DR17", ``AEGIS", ``SEGUE", ``Gaia", ``Gaia-ESO", \\
$-$ & ``LAMOST", ``RAVE", ``LIT", and ``BB" to indicate the source of radial velocity from \citet{Huang2022}, \\
$-$ & and ``GaiaDR3", ``LAMOST", ``SEGUE", ``APOGEE", ``GALAH", and ``RAVE" from \citet{Huang2023} [rv\_adop\_flg] & $-$\\
plx & The parallax from Gaia DR3 [parallax] & mas\\
e\_plx & The parallax uncertainty from Gaia DR3 [parallax\_error] & mas\\
pmRA & The proper motion in the Right Ascension from Gaia DR3 [pmra] & mas yr$^{-1}$\\
e\_pmRA & The proper motion uncertainty in the Right Ascension from Gaia DR3 [pmra\_error] & mas yr$^{-1}$\\
pmDE &	The proper motion in the Declination from Gaia DR3 [pmdec] & mas yr$^{-1}$\\
e\_pmDE & The proper motion uncertainty in the Declination from Gaia DR3 [pmdec\_error] & mas yr$^{-1}$\\
pmRApmDEcor & The correlation coefficient between the proper motion in Right Ascension and in Declination from Gaia DR3 & $-$\\
Type & Flag with ``Dwarf" and ``Giant" from \citet{Huang2022, Huang2023} & $-$\\
SubType & Flag with ``TO" for turn-off stars and ``MS" for main-sequence stars from \citet{Huang2022, Huang2023} & $-$\\
vPHI & The rotational velocity as given by \texttt{\texttt{AGAMA}} & km s$^{-1}$\\
e\_vPHI & The rotational velocity uncertainty as given by Monte Carlo sampling through \texttt{AGAMA} & km s$^{-1}$\\
E & The orbital energy as given by \texttt{AGAMA} & km$^{2}$ s$^{-2}$\\
e\_E & The orbital energy uncertainty as given by \texttt{AGAMA} & km$^{2}$ s$^{-2}$\\
Jr,Jphi,Jz & The cylindrical actions as given by \texttt{AGAMA} & kpc km s$^{-1}$\\
e\_Jr,e\_Jphi,e\_Jz & The cylindrical actions uncertainty as given by \texttt{AGAMA} & kpc km s$^{-1}$\\
rperi & The Galactic pericentric distance as given by \texttt{AGAMA} & kpc\\
e\_rperi & The Galactic pericentric distance uncertainty as given by \texttt{AGAMA} & kpc\\
rapo & The Galactic apocentric distance as given by \texttt{AGAMA} & kpc\\
e\_rapo & The Galactic apocentric distance uncertainty as given by Monte Carlo sampling through \texttt{AGAMA} & kpc\\
Zmax & The maximum height above the Galactic plane as given by \texttt{AGAMA} & kpc\\
e\_Zmax & The maximum height uncertainty above the Galactic plane as given by Monte Carlo sampling through \texttt{AGAMA} & kpc\\
relerr-Zmax & The relative uncertainty of the maximum height above the Galactic plane & $-$\\
Rmax & The projection of the Galactic apocentric distance onto the Galactic plane as given by $\sqrt{r_{\rm apo}^2 - Z_{\rm max}^2}$ & kpc\\
relerr-Rmax & The relative uncertainty of the projection of the Galactic apocentric distance onto the Galactic plane. & \\
$-$ & The uncertainty is as given by Monte Carlo sampling through \texttt{AGAMA} & $-$\\
IA & The inclination angle defined as the arctangent ratio of ($Z_{\rm max}$ / $R_{\rm max}$) & rad\\
ecc & The eccentricity as given by ($r_{\rm apo}$ - $r_{\rm peri}$) / ($r_{\rm apo}$ + $r_{\rm peri}$) through \texttt{AGAMA} & $-$\\
criterion & Flag with ``Z$_{\rm max}$", ``Haywood", or ``Both" to indicate the criterion used for the separation of stars on halo-like orbits from\\
$-$ & those on thick-disk like and thin-disk like orbits & $-$
\enddata
\end{deluxetable*}

%% file: finalcatalog_stub.tex
\startlongtable
\begin{deluxetable}{ccccccccccccccc}
\tabletypesize{\scriptsize}
\tablewidth{0pt}
\tablecaption{\label{placeholder}}
\tablecaption{1,496 VMP/EMP Disk-like Candidate Stars in the Combined SMSS/SAGES Sample \label{tab:finalcatalog}}
\tablehead{\colhead{Source ID} & \colhead{R.A.} & \colhead{Decl.} & \colhead{\textit{G$_{C}$}} & \colhead{$G_{\rm BP} - G_{\rm RP}$ }& \colhead{Type} & \colhead{Dist} & \colhead{[Fe/H]} & \colhead{err$_{\rm [Fe/H]}$} & \colhead{v$_{\phi}$} & \colhead{$Z_{\rm max}$} & \colhead{$R_{\rm max}$} & \colhead{$IA$} & \colhead{ecc} & \colhead{criterion} 
    \\[-2ex] 
    \colhead{} & \colhead{(J2000)} & \colhead{(J2000)} & \colhead{} & \colhead{} & \colhead{} & \colhead{(kpc)} & \colhead{} & \colhead{(dex)} & \colhead{(km s$^{-1}$)} & \colhead{(kpc)} & \colhead{(kpc)} & \colhead{(rad)} & \colhead{} & \colhead{}}
\startdata
4918300611248859648 & 00:00:09.07 & -59:03:26.0 & 14.67 & 0.89 & G & 2.48 & $-$2.10 & 0.15 & 179.3 & 2.5 & 7.2 & 0.33 & 0.175 & Both\\
2853059333495692800 & 00:00:18.21 & +25:05:15.2 & 17.99 & 1.08 & D & 1.65 & $-$2.01 & 0.33 & 160.3 & 1.3 & 8.9 & 0.14 & 0.317 & Both\\
2853263288606167808 & 00:00:58.03 & +26:00:23.4 & 17.98 & 0.99 & D & 2.76 & $-$2.21 & 0.33 & 170.5 & 2.1 & 9.4 & 0.22 & 0.271 & Both\\
2880219431068389632 & 00:02:39.84 & +36:50:18.5 & 17.73 & 0.87 & D & 2.14 & $-$2.06 & 0.35 & 224.8 & 1.1 & 10.1 & 0.11 & 0.125 & Both\\
2421349012110147840 & 00:03:51.14 & -12:31:54.1 & 12.65 & 1.03 & G & 2.77 & $-$2.24 & 0.06 & 181.5 & 3.4 & 8.9 & 0.36 & 0.252 & Haywood\\
2876918628442545152 & 00:06:55.97 & +35:54:29.0 & 17.50 & 0.84 & D & 2.18 & $-$2.70 & 0.21 & 190.7 & 1.0 & 9.1 & 0.10 & 0.171 & Both\\
2546602108275485568 & 00:06:57.29 & +01:15:53.4 & 17.26 & 0.72 & D & 1.83 & $-$2.66 & 0.20 & 158.5 & 1.8 & 9.4 & 0.19 & 0.384 & Both\\
2319766674712899840 & 00:07:42.57 & -30:56:13.4 & 11.72 & 0.96 & G & 1.57 & $-$2.08 & 0.04 & 179.0 & 2.0 & 8.2 & 0.23 & 0.240 & Both\\
2739001632226213504 & 00:08:36.17 & +02:34:11.1 & 17.26 & 0.70 & D & 1.86 & $-$2.53 & 0.21 & 162.1 & 5.9 & 12.6 & 0.44 & 0.513 & Haywood\\
2443519530213095168 & 00:09:47.75 & -06:15:31.9 & 17.60 & 1.01 & D & 1.66 & $-$2.16 & 0.14 & 174.9 & 1.9 & 8.4 & 0.23 & 0.238 & Both\\
2546356509161321600 & 00:11:08.63 & +00:59:50.1 & 15.42 & 0.73 & D & 1.06 & $-$2.12 & 0.08 & 151.5 & 1.1 & 10.5 & 0.10 & 0.491 & Both\\
2740672133986589568 & 00:12:26.18 & +03:34:41.9 & 16.52 & 0.63 & D & 1.79 & $-$2.21 & 0.19 & 160.3 & 2.0 & 8.9 & 0.22 & 0.333 & Both\\
2850385316920406016 & 00:14:20.01 & +25:45:11.1 & 17.15 & 1.05 & D & 1.17 & $-$2.12 & 0.19 & 195.2 & 1.3 & 9.3 & 0.14 & 0.218 & Both\\
395035668663587328 & 00:15:24.23 & +51:39:07.5 & 13.23 & 0.68 & D & 0.78 & $-$3.18 & 0.08 & 239.3 & 0.2 & 9.1 & 0.02 & 0.030 & Both\\
2768462152939050112 & 00:17:21.85 & +14:49:24.1 & 16.57 & 0.62 & D & 1.40 & $-$2.10 & 0.21 & 167.2 & 1.2 & 10.0 & 0.12 & 0.387 & Both\\
2428249340927211008 & 00:18:20.37 & -09:18:32.9 & 15.43 & 0.64 & D & 1.24 & $-$2.66 & 0.36 & 165.3 & 2.1 & 9.4 & 0.22 & 0.372 & Both\\
2800378157994808576 & 00:18:48.76 & +22:15:30.1 & 16.28 & 0.80 & D & 1.17 & $-$2.37 & 0.12 & 195.0 & 4.6 & 18.1 & 0.25 & 0.609 & Haywood\\
2876157251000218496 & 00:18:56.93 & +34:54:44.3 & 17.12 & 0.79 & D & 2.03 & $-$2.84 & 0.27 & 231.4 & 1.0 & 10.3 & 0.10 & 0.111 & Both\\
4991862612070916736 & 00:22:18.35 & -43:31:15.4 & 14.36 & 0.82 & G & 2.13 & $-$2.68 & 0.30 & 185.0 & 4.3 & 8.2 & 0.49 & 0.197 & Haywood\\
2864157352892856576 & 00:22:27.17 & +34:34:14.0 & 18.43 & 0.85 & D & 3.30 & $-$2.68 & 0.35 & 155.2 & 2.8 & 12.6 & 0.22 & 0.491 & Both\\
378314738205895296 & 00:23:19.23 & +36:02:21.5 & 17.73 & 0.86 & D & 2.29 & $-$2.12 & 0.30 & 188.3 & 4.2 & 9.5 & 0.41 & 0.165 & Haywood\\
2747658873250089600 & 00:23:27.30 & +06:25:53.5 & 17.40 & 0.87 & D & 1.63 & $-$2.23 & 0.20 & 210.7 & 16.6 & 25.0 & 0.58 & 0.692 & Haywood\\
2855523884512993792 & 00:24:08.92 & +26:09:43.0 & 17.67 & 0.74 & D & 2.53 & $-$2.06 & 0.33 & 220.7 & 1.6 & 9.4 & 0.17 & 0.055 & Both\\
2367276232846933760 & 00:26:35.09 & -17:49:21.4 & 16.62 & 0.62 & D & 2.67 & $-$2.28 & 0.39 & 231.7 & 6.5 & 10.7 & 0.55 & 0.195 & Haywood\\
2800046311641547392 & 00:26:49.64 & +22:30:39.2 & 14.57 & 0.95 & G & 5.83 & $-$2.62 & 0.09 & 185.4 & 6.4 & 10.9 & 0.53 & 0.161 & Haywood\\
2426849250308221696 & 00:28:21.09 & -08:46:37.8 & 13.48 & 0.98 & G & 3.77 & $-$2.31 & 0.06 & 161.2 & 6.4 & 9.7 & 0.58 & 0.342 & Haywood\\
4918673723647798528 & 00:30:35.34 & -57:32:26.5 & 13.88 & 0.94 & G & 3.57 & $-$2.22 & 0.12 & 180.5 & 3.2 & 7.6 & 0.40 & 0.215 & Haywood\\
2806067477833582336 & 00:32:20.46 & +23:14:27.8 & 13.24 & 0.67 & D & 0.42 & $-$2.75 & 0.03 & 222.7 & 0.3 & 10.3 & 0.03 & 0.232 & Both\\
2542770623787256576 & 00:34:51.59 & -00:28:22.3 & 17.45 & 1.00 & D & 1.58 & $-$2.20 & 0.20 & 250.2 & 2.0 & 11.3 & 0.18 & 0.181 & Both\\
2750377553188528512 & 00:35:21.56 & +09:37:58.8 & 17.54 & 0.69 & D & 2.10 & $-$2.22 & 0.29 & 167.2 & 2.6 & 12.1 & 0.21 & 0.481 & Both\\
2425850893750224000 & 00:38:34.18 & -10:03:57.2 & 16.61 & 0.66 & D & 2.18 & $-$2.68 & 0.44 & 296.2 & 4.6 & 17.0 & 0.26 & 0.359 & Haywood\\
2781291215956696704 & 00:43:40.61 & +15:30:11.6 & 17.54 & 0.88 & D & 1.68 & $-$2.14 & 0.27 & 176.3 & 1.5 & 10.7 & 0.14 & 0.379 & Both\\
2776361869106548992 & 00:45:00.93 & +13:45:45.6 & 16.95 & 0.82 & D & 0.88 & $-$2.28 & 0.17 & 238.8 & 10.2 & 14.3 & 0.62 & 0.359 & Haywood\\
2530949495102121600 & 00:46:02.00 & -00:26:48.1 & 16.65 & 0.72 & D & 1.74 & $-$2.24 & 0.09 & 164.1 & 1.8 & 9.7 & 0.18 & 0.364 & Both\\
2776181789717803008 & 00:48:25.66 & +13:41:47.4 & 13.84 & 0.84 & G & 2.27 & $-$2.50 & 0.40 & 191.3 & 2.5 & 12.7 & 0.19 & 0.409 & Both\\
4999763977506992256 & 00:55:17.54 & -38:28:37.7 & 14.23 & 0.64 & D & 0.94 & $-$2.42 & 0.23 & 202.9 & 1.0 & 8.7 & 0.11 & 0.181 & Both\\
378064947205609984 & 00:55:28.52 & +46:24:06.5 & 13.77 & 1.09 & G & 0.86 & $-$2.03 & 0.03 & 151.6 & 0.6 & 8.8 & 0.07 & 0.357 & Both\\
364777486620366976 & 00:55:46.05 & +38:18:03.6 & 17.28 & 0.64 & D & 1.96 & $-$2.07 & 0.35 & 198.6 & 1.0 & 9.6 & 0.11 & 0.172 & Both\\
2357297065313248512 & 00:57:51.87 & -18:12:25.9 & 12.74 & 0.62 & D & 0.44 & $-$2.54 & 0.15 & 185.3 & 0.9 & 8.7 & 0.10 & 0.239 & Both\\
2553447805108682496 & 00:58:27.10 & +06:33:56.4 & 12.25 & 1.08 & G & 3.51 & $-$2.62 & 0.02 & 165.1 & 3.2 & 11.0 & 0.29 & 0.367 & Haywood\\
374902167646414720 & 01:00:04.74 & +43:02:08.5 & 15.11 & 1.19 & G & 18.53 & $-$2.29 & 0.07 & 182.7 & 14.2 & 27.7 & 0.47 & 0.311 & Haywood\\
4710497998840333568 & 01:01:03.19 & -62:17:20.8 & 13.48 & 0.97 & G & 3.61 & $-$2.50 & 0.07 & 234.9 & 8.8 & 16.5 & 0.49 & 0.537 & Haywood\\
4684461048100616448 & 01:01:30.40 & -75:44:51.0 & 14.04 & 1.01 & G & 7.28 & $-$2.13 & 0.05 & 156.3 & 5.6 & 8.3 & 0.60 & 0.371 & Haywood\\
2536647615329253376 & 01:02:09.79 & +01:03:49.3 & 15.92 & 0.72 & D & 0.40 & $-$2.34 & 0.09 & 172.9 & 0.5 & 8.6 & 0.06 & 0.279 & Both\\
2788522394695432704 & 01:02:50.31 & +19:26:45.8 & 17.58 & 0.65 & D & 2.09 & $-$2.41 & 0.29 & 235.7 & 1.7 & 10.6 & 0.16 & 0.110 & Both\\
2536576662469536768 & 01:03:30.39 & +00:43:47.5 & 17.13 & 0.61 & D & 1.44 & $-$2.70 & 0.29 & 187.7 & 1.3 & 8.6 & 0.15 & 0.181 & Both\\
4926978266253100672 & 01:03:32.61 & -53:46:54.4 & 13.98 & 0.95 & G & 4.17 & $-$2.35 & 0.10 & 203.4 & 5.9 & 8.4 & 0.61 & 0.164 & Haywood\\
2538090861779751424 & 01:04:23.30 & +01:00:00.8 & 17.73 & 0.70 & D & 2.17 & $-$2.73 & 0.24 & 155.9 & 2.6 & 9.4 & 0.27 & 0.346 & Both\\
369849739917996544 & 01:07:30.66 & +38:22:04.8 & 17.23 & 0.70 & D & 1.45 & $-$3.14 & 0.35 & 224.5 & 0.7 & 9.1 & 0.07 & 0.044 & Both\\
\enddata
\tablecomments{This table is a stub; the full table is available in the electronic edition.}
\end{deluxetable}

%% file: main.bbl
\begin{thebibliography}{}
\expandafter\ifx\csname natexlab\endcsname\relax\def\natexlab#1{#1}\fi
\providecommand{\url}[1]{\href{#1}{#1}}
\providecommand{\dodoi}[1]{doi:~\href{http://doi.org/#1}{\nolinkurl{#1}}}
\providecommand{\doeprint}[1]{\href{http://ascl.net/#1}{\nolinkurl{http://ascl.net/#1}}}
\providecommand{\doarXiv}[1]{\href{https://arxiv.org/abs/#1}{\nolinkurl{https://arxiv.org/abs/#1}}}

\bibitem[{{An} \& {Beers}(2020)}]{An2020}
{An}, D., \& {Beers}, T.~C. 2020, \apj, 897, 39, \dodoi{10.3847/1538-4357/ab8d39}

\bibitem[{{An} \& {Beers}(2021{\natexlab{a}})}]{An2021a}
---. 2021{\natexlab{a}}, \apj, 907, 101, \dodoi{10.3847/1538-4357/abccd2}

\bibitem[{{An} \& {Beers}(2021{\natexlab{b}})}]{An2021b}
---. 2021{\natexlab{b}}, \apj, 918, 74, \dodoi{10.3847/1538-4357/ac07a4}

\bibitem[{{An} {et~al.}(2023){An}, {Beers}, {Lee}, \& {Masseron}}]{An2023}
{An}, D., {Beers}, T.~C., {Lee}, Y.~S., \& {Masseron}, T. 2023, \apj, 952, 66, \dodoi{10.3847/1538-4357/acd5cb}

\bibitem[{{Bailer-Jones} {et~al.}(2021){Bailer-Jones}, {Rybizki}, {Fouesneau}, {Demleitner}, \& {Andrae}}]{Bailer-Jones2021}
{Bailer-Jones}, C.~A.~L., {Rybizki}, J., {Fouesneau}, M., {Demleitner}, M., \& {Andrae}, R. 2021, \aj, 161, 147, \dodoi{10.3847/1538-3881/abd806}

\bibitem[{{Baumgardt} \& {Vasiliev}(2021)}]{Baumgardt2021}
{Baumgardt}, H., \& {Vasiliev}, E. 2021, \mnras, 505, 5957, \dodoi{10.1093/mnras/stab1474}

\bibitem[{{Beers} \& {Christlieb}(2005)}]{Beers2005}
{Beers}, T.~C., \& {Christlieb}, N. 2005, \araa, 43, 531, \dodoi{10.1146/annurev.astro.42.053102.134057}

\bibitem[{{Beers} {et~al.}(1990){Beers}, {Flynn}, \& {Gebhardt}}]{Beers1990a}
{Beers}, T.~C., {Flynn}, K., \& {Gebhardt}, K. 1990, \aj, 100, 32, \dodoi{10.1086/115487}

\bibitem[{{Beers} {et~al.}(2014){Beers}, {Norris}, {Placco}, {Lee}, {Rossi}, {Carollo}, \& {Masseron}}]{Beers2014}
{Beers}, T.~C., {Norris}, J.~E., {Placco}, V.~M., {et~al.} 2014, \apj, 794, 58, \dodoi{10.1088/0004-637X/794/1/58}

\bibitem[{{Beers} {et~al.}(1985){Beers}, {Preston}, \& {Shectman}}]{Beers1985}
{Beers}, T.~C., {Preston}, G.~W., \& {Shectman}, S.~A. 1985, \aj, 90, 2089, \dodoi{10.1086/113917}

\bibitem[{{Beers} {et~al.}(1992){Beers}, {Preston}, \& {Shectman}}]{Beers1992}
---. 1992, \aj, 103, 1987, \dodoi{10.1086/116207}

\bibitem[{{Bellazzini} {et~al.}(2024){Bellazzini}, {Massari}, {Ceccarelli}, {Mucciarelli}, {Bragaglia}, {Riello}, {De Angeli}, \& {Montegriffo}}]{Bellazzini2024}
{Bellazzini}, M., {Massari}, D., {Ceccarelli}, E., {et~al.} 2024, \aap, 683, A136, \dodoi{10.1051/0004-6361/202348106}

\bibitem[{{Bellazzini} {et~al.}(2023){Bellazzini}, {Massari}, {De Angeli}, {Mucciarelli}, {Bragaglia}, {Riello}, \& {Montegriffo}}]{Bellazzini2023}
{Bellazzini}, M., {Massari}, D., {De Angeli}, F., {et~al.} 2023, \aap, 674, A194, \dodoi{10.1051/0004-6361/202345921}

\bibitem[{{Belokurov} {et~al.}(2018){Belokurov}, {Erkal}, {Evans}, {Koposov}, \& {Deason}}]{Belokurov2018}
{Belokurov}, V., {Erkal}, D., {Evans}, N.~W., {Koposov}, S.~E., \& {Deason}, A.~J. 2018, \mnras, 478, 611, \dodoi{10.1093/mnras/sty982}

\bibitem[{{Bressan} {et~al.}(2012){Bressan}, {Marigo}, {Girardi}, {Salasnich}, {Dal Cero}, {Rubele}, \& {Nanni}}]{Bressan2012}
{Bressan}, A., {Marigo}, P., {Girardi}, L., {et~al.} 2012, \mnras, 427, 127, \dodoi{10.1111/j.1365-2966.2012.21948.x}

\bibitem[{{Carollo} {et~al.}(2010){Carollo}, {Beers}, {Chiba}, {Norris}, {Freeman}, \& {Lee}}]{Carollo2010}
{Carollo}, D., {Beers}, T.~C., {Chiba}, M., {et~al.} 2010, in Chemical Abundances in the Universe: Connecting First Stars to Planets, ed. K.~{Cunha}, M.~{Spite}, \& B.~{Barbuy}, Vol. 265, 267--270, \dodoi{10.1017/S1743921310000724}

\bibitem[{{Carollo} {et~al.}(2023){Carollo}, {Christlieb}, {Tissera}, \& {Sillero}}]{Carollo2023}
{Carollo}, D., {Christlieb}, N., {Tissera}, P.~B., \& {Sillero}, E. 2023, \apj, 946, 99, \dodoi{10.3847/1538-4357/acac25}

\bibitem[{{Carollo} {et~al.}(2007){Carollo}, {Beers}, {Lee}, {Chiba}, {Norris}, {Wilhelm}, {Sivarani}, {Marsteller}, {Munn}, {Bailer-Jones}, {Fiorentin}, \& {York}}]{carollo2007}
{Carollo}, D., {Beers}, T.~C., {Lee}, Y.~S., {et~al.} 2007, \nat, 450, 1020, \dodoi{10.1038/nature06460}

\bibitem[{{Carollo} {et~al.}(2019){Carollo}, {Chiba}, {Ishigaki}, {Freeman}, {Beers}, {Lee}, {Tissera}, {Battistini}, \& {Primas}}]{Carollo2019}
{Carollo}, D., {Chiba}, M., {Ishigaki}, M., {et~al.} 2019, \apj, 887, 22, \dodoi{10.3847/1538-4357/ab517c}

\bibitem[{{Carter} {et~al.}(2021){Carter}, {Conroy}, {Zaritsky}, {Ting}, {Bonaca}, {Naidu}, {Johnson}, {Cargile}, {Caldwell}, {Speagle}, \& {Han}}]{Carter2021}
{Carter}, C., {Conroy}, C., {Zaritsky}, D., {et~al.} 2021, \apj, 908, 208, \dodoi{10.3847/1538-4357/abcda4}

\bibitem[{{Cenarro} {et~al.}(2019){Cenarro}, {Moles}, {Crist{\'o}bal-Hornillos}, {Mar{\'\i}n-Franch}, {Ederoclite}, {Varela}, {L{\'o}pez-Sanjuan}, {Hern{\'a}ndez-Monteagudo}, {Angulo}, {V{\'a}zquez Rami{\'o}}, {Viironen}, {Bonoli}, {Orsi}, {Hurier}, {San Roman}, {Greisel}, {Vilella-Rojo}, {D{\'\i}az-Garc{\'\i}a}, {Logro{\~n}o-Garc{\'\i}a}, {Gurung-L{\'o}pez}, {Spinoso}, {Izquierdo-Villalba}, {Aguerri}, {Allende Prieto}, {Bonatto}, {Carvano}, {Chies-Santos}, {Daflon}, {Dupke}, {Falc{\'o}n-Barroso}, {Gon{\c{c}}alves}, {Jim{\'e}nez-Teja}, {Molino}, {Placco}, {Solano}, {Whitten}, {Abril}, {Ant{\'o}n}, {Bello}, {Bielsa de Toledo}, {Castillo-Ram{\'\i}rez}, {Chueca}, {Civera}, {D{\'\i}az-Mart{\'\i}n}, {Dom{\'\i}nguez-Mart{\'\i}nez}, {Garzar{\'a}n-Calderaro}, {Hern{\'a}ndez-Fuertes}, {Iglesias-Marzoa}, {I{\~n}iguez}, {Jim{\'e}nez Ruiz}, {Kruuse}, {Lamadrid}, {Lasso-Cabrera}, {L{\'o}pez-Alegre}, {L{\'o}pez-Sainz}, {Ma{\'\i}cas}, {Moreno-Signes}, {Muniesa}, {Rodr{\'\i}guez-Llano}, {Rueda-Teruel}, {Rueda-Teruel},
  {Soriano-Lagu{\'\i}a}, {Tilve}, {Valdivielso}, {Yanes-D{\'\i}az}, {Alcaniz}, {Mendes de Oliveira}, {Sodr{\'e}}, {Coelho}, {Lopes de Oliveira}, {Tamm}, {Xavier}, {Abramo}, {Akras}, {Alfaro}, {Alvarez-Candal}, {Ascaso}, {Beasley}, {Beers}, {Borges Fernandes}, {Bruzual}, {Buzzo}, {Carrasco}, {Cepa}, {Cortesi}, {Costa-Duarte}, {De Pr{\'a}}, {Favole}, {Galarza}, {Galbany}, {Garcia}, {Gonz{\'a}lez Delgado}, {Gonz{\'a}lez-Serrano}, {Guti{\'e}rrez-Soto}, {Hernandez-Jimenez}, {Kanaan}, {Kuncarayakti}, {Landim}, {Laur}, {Licandro}, {Lima Neto}, {Lyman}, {Ma{\'\i}z Apell{\'a}niz}, {Miralda-Escud{\'e}}, {Morate}, {Nogueira-Cavalcante}, {Novais}, {Oncins}, {Oteo}, {Overzier}, {Pereira}, {Rebassa-Mansergas}, {Reis}, {Roig}, {Sako}, {Salvador-Rusi{\~n}ol}, {Sampedro}, {S{\'a}nchez-Bl{\'a}zquez}, {Santos}, {Schmidtobreick}, {Siffert}, {Telles}, \& {Vilchez}}]{Cenarro2019}
{Cenarro}, A.~J., {Moles}, M., {Crist{\'o}bal-Hornillos}, D., {et~al.} 2019, \aap, 622, A176, \dodoi{10.1051/0004-6361/201833036}

\bibitem[{{Chambers} {et~al.}(2016){Chambers}, {Magnier}, {Metcalfe}, {Flewelling}, {Huber}, {Waters}, {Denneau}, {Draper}, {Farrow}, {Finkbeiner}, {Holmberg}, {Koppenhoefer}, {Price}, {Rest}, {Saglia}, {Schlafly}, {Smartt}, {Sweeney}, {Wainscoat}, {Burgett}, {Chastel}, {Grav}, {Heasley}, {Hodapp}, {Jedicke}, {Kaiser}, {Kudritzki}, {Luppino}, {Lupton}, {Monet}, {Morgan}, {Onaka}, {Shiao}, {Stubbs}, {Tonry}, {White}, {Ba{\~n}ados}, {Bell}, {Bender}, {Bernard}, {Boegner}, {Boffi}, {Botticella}, {Calamida}, {Casertano}, {Chen}, {Chen}, {Cole}, {Deacon}, {Frenk}, {Fitzsimmons}, {Gezari}, {Gibbs}, {Goessl}, {Goggia}, {Gourgue}, {Goldman}, {Grant}, {Grebel}, {Hambly}, {Hasinger}, {Heavens}, {Heckman}, {Henderson}, {Henning}, {Holman}, {Hopp}, {Ip}, {Isani}, {Jackson}, {Keyes}, {Koekemoer}, {Kotak}, {Le}, {Liska}, {Long}, {Lucey}, {Liu}, {Martin}, {Masci}, {McLean}, {Mindel}, {Misra}, {Morganson}, {Murphy}, {Obaika}, {Narayan}, {Nieto-Santisteban}, {Norberg}, {Peacock}, {Pier}, {Postman}, {Primak}, {Rae}, {Rai},
  {Riess}, {Riffeser}, {Rix}, {R{\"o}ser}, {Russel}, {Rutz}, {Schilbach}, {Schultz}, {Scolnic}, {Strolger}, {Szalay}, {Seitz}, {Small}, {Smith}, {Soderblom}, {Taylor}, {Thomson}, {Taylor}, {Thakar}, {Thiel}, {Thilker}, {Unger}, {Urata}, {Valenti}, {Wagner}, {Walder}, {Walter}, {Watters}, {Werner}, {Wood-Vasey}, \& {Wyse}}]{Chambers2016}
{Chambers}, K.~C., {Magnier}, E.~A., {Metcalfe}, N., {et~al.} 2016, arXiv e-prints, arXiv:1612.05560, \dodoi{10.48550/arXiv.1612.05560}

\bibitem[{{Christlieb}(2003)}]{Christlieb2003}
{Christlieb}, N. 2003, Reviews in Modern Astronomy, 16, 191, \dodoi{10.1002/9783527617647.ch8}

\bibitem[{{Conroy} {et~al.}(2019){Conroy}, {Naidu}, {Zaritsky}, {Bonaca}, {Cargile}, {Johnson}, \& {Caldwell}}]{Conroy2019}
{Conroy}, C., {Naidu}, R.~P., {Zaritsky}, D., {et~al.} 2019, \apj, 887, 237, \dodoi{10.3847/1538-4357/ab5710}

\bibitem[{{Cordoni} {et~al.}(2021){Cordoni}, {Da Costa}, {Yong}, {Mackey}, {Marino}, {Monty}, {Nordlander}, {Norris}, {Asplund}, {Bessell}, {Casey}, {Frebel}, {Lind}, {Murphy}, {Schmidt}, {Gao}, {Xylakis-Dornbusch}, {Amarsi}, \& {Milone}}]{Cordoni2021}
{Cordoni}, G., {Da Costa}, G.~S., {Yong}, D., {et~al.} 2021, \mnras, 503, 2539, \dodoi{10.1093/mnras/staa3417}

\bibitem[{{De Silva} {et~al.}(2015){De Silva}, {Freeman}, {Bland-Hawthorn}, {Martell}, {de Boer}, {Asplund}, {Keller}, {Sharma}, {Zucker}, {Zwitter}, {Anguiano}, {Bacigalupo}, {Bayliss}, {Beavis}, {Bergemann}, {Campbell}, {Cannon}, {Carollo}, {Casagrande}, {Casey}, {Da Costa}, {D'Orazi}, {Dotter}, {Duong}, {Heger}, {Ireland}, {Kafle}, {Kos}, {Lattanzio}, {Lewis}, {Lin}, {Lind}, {Munari}, {Nataf}, {O'Toole}, {Parker}, {Reid}, {Schlesinger}, {Sheinis}, {Simpson}, {Stello}, {Ting}, {Traven}, {Watson}, {Wittenmyer}, {Yong}, \& {{\v{Z}}erjal}}]{DeSilva2015}
{De Silva}, G.~M., {Freeman}, K.~C., {Bland-Hawthorn}, J., {et~al.} 2015, \mnras, 449, 2604, \dodoi{10.1093/mnras/stv327}

\bibitem[{{Deng} {et~al.}(2012){Deng}, {Newberg}, {Liu}, {Carlin}, {Beers}, {Chen}, {Chen}, {Christlieb}, {Grillmair}, {Guhathakurta}, {Han}, {Hou}, {Lee}, {L{\'e}pine}, {Li}, {Liu}, {Pan}, {Sellwood}, {Wang}, {Wang}, {Yang}, {Yanny}, {Zhang}, {Zhang}, {Zheng}, \& {Zhu}}]{Deng2012}
{Deng}, L.-C., {Newberg}, H.~J., {Liu}, C., {et~al.} 2012, Research in Astronomy and Astrophysics, 12, 735, \dodoi{10.1088/1674-4527/12/7/003}

\bibitem[{{Di Matteo} {et~al.}(2020){Di Matteo}, {Spite}, {Haywood}, {Bonifacio}, {G{\'o}mez}, {Spite}, \& {Caffau}}]{DiMatteo2020}
{Di Matteo}, P., {Spite}, M., {Haywood}, M., {et~al.} 2020, \aap, 636, A115, \dodoi{10.1051/0004-6361/201937016}

\bibitem[{{Fan} {et~al.}(2023){Fan}, {Zhao}, {Wang}, {Zheng}, {Zhao}, {Li}, {Chen}, {Yuan}, {Li}, {Tan}, {Song}, {Zuo}, {Huang}, {Luo}, {Esamdin}, {Ma}, {Li}, {Song}, {Grupp}, {Zhao}, {Ehgamberdiev}, {Burkhonov}, {Feng}, {Bai}, {Zhang}, {Niu}, {Khodjaev}, {Khafizov}, {Asfandiyarov}, {Shaymanov}, {Karimov}, {Yuldashev}, {Lu}, {Zhaori}, {Hong}, {Hu}, {Liu}, \& {Xu}}]{Fan2023}
{Fan}, Z., {Zhao}, G., {Wang}, W., {et~al.} 2023, \apjs, 268, 9, \dodoi{10.3847/1538-4365/ace04a}

\bibitem[{{Feltzing} \& {Feuillet}(2023)}]{Feltzing2023}
{Feltzing}, S., \& {Feuillet}, D. 2023, \apj, 953, 143, \dodoi{10.3847/1538-4357/ace185}

\bibitem[{{Fern{\'a}ndez-Alvar} {et~al.}(2021){Fern{\'a}ndez-Alvar}, {Kordopatis}, {Hill}, {Starkenburg}, {Viswanathan}, {Martin}, {Thomas}, {Navarro}, {Malhan}, {Sestito}, {Gonz{\'a}lez Hern{\'a}ndez}, \& {Carlberg}}]{Fernandez-Alvar2021}
{Fern{\'a}ndez-Alvar}, E., {Kordopatis}, G., {Hill}, V., {et~al.} 2021, \mnras, 508, 1509, \dodoi{10.1093/mnras/stab2617}

\bibitem[{{Gaia Collaboration} {et~al.}(2023){Gaia Collaboration}, {Montegriffo}, {Bellazzini}, {De Angeli}, {Andrae}, {Barstow}, {Bossini}, {Bragaglia}, {Burgess}, {Cacciari}, {Carrasco}, {Chornay}, {Delchambre}, {Evans}, {Fouesneau}, {Fr{\'e}mat}, {Garabato}, {Jordi}, {Manteiga}, {Massari}, {Palaversa}, {Pancino}, {Riello}, {Ruz Mieres}, {Sanna}, {Santove{\~n}a}, {Sordo}, {Vallenari}, {Walton}, {Brown}, {Prusti}, {de Bruijne}, {Arenou}, {Babusiaux}, {Biermann}, {Creevey}, {Ducourant}, {Eyer}, {Guerra}, {Hutton}, {Klioner}, {Lammers}, {Lindegren}, {Luri}, {Mignard}, {Panem}, {Pourbaix}, {Randich}, {Sartoretti}, {Soubiran}, {Tanga}, {Bailer-Jones}, {Bastian}, {Drimmel}, {Jansen}, {Katz}, {Lattanzi}, {van Leeuwen}, {Bakker}, {Casta{\~n}eda}, {Fabricius}, {Galluccio}, {Guerrier}, {Heiter}, {Masana}, {Messineo}, {Mowlavi}, {Nicolas}, {Nienartowicz}, {Pailler}, {Panuzzo}, {Riclet}, {Roux}, {Seabroke}, {Th{\'e}venin}, {Gracia-Abril}, {Portell}, {Teyssier}, {Altmann}, {Audard}, {Bellas-Velidis}, {Benson},
  {Berthier}, {Blomme}, {Busonero}, {Busso}, {C{\'a}novas}, {Carry}, {Cellino}, {Cheek}, {Clementini}, {Damerdji}, {Davidson}, {de Teodoro}, {Nu{\~n}ez Campos}, {Dell'Oro}, {Esquej}, {Fern{\'a}ndez-Hern{\'a}ndez}, {Fraile}, {Garc{\'\i}a-Lario}, {Gosset}, {Haigron}, {Halbwachs}, {Hambly}, {Harrison}, {Hern{\'a}ndez}, {Hestroffer}, {Hodgkin}, {Holl}, {Jan{\ss}en}, {Jevardat de Fombelle}, {Jordan}, {Krone-Martins}, {Lanzafame}, {L{\"o}ffler}, {Marchal}, {Marrese}, {Moitinho}, {Muinonen}, {Osborne}, {Pauwels}, {Recio-Blanco}, {Reyl{\'e}}, {Rimoldini}, {Roegiers}, {Rybizki}, {Sarro}, {Siopis}, {Smith}, {Sozzetti}, {Utrilla}, {van Leeuwen}, {Abbas}, {{\'A}brah{\'a}m}, {Abreu Aramburu}, {Aerts}, {Aguado}, {Ajaj}, {Aldea-Montero}, {Altavilla}, {{\'A}lvarez}, {Alves}, {Anderson}, {Anglada Varela}, {Antoja}, {Baines}, {Baker}, {Balaguer-N{\'u}{\~n}ez}, {Balbinot}, {Balog}, {Barache}, {Barbato}, {Barros}, {Bartolom{\'e}}, {Bassilana}, {Bauchet}, {Becciani}, {Berihuete}, {Bernet}, {Bertone}, {Bianchi}, {Binnenfeld},
  {Blanco-Cuaresma}, {Boch}, {Bombrun}, {Bouquillon}, {Bramante}, {Breedt}, {Bressan}, {Brouillet}, {Brugaletta}, {Bucciarelli}, {Burlacu}, {Butkevich}, {Buzzi}, {Caffau}, {Cancelliere}, {Cantat-Gaudin}, {Carballo}, {Carlucci}, {Carnerero}, {Casamiquela}, {Castellani}, {Castro-Ginard}, {Chaoul}, {Charlot}, {Chemin}, {Chiaramida}, {Chiavassa}, {Comoretto}, {Contursi}, {Cooper}, {Cornez}, {Cowell}, {Crifo}, {Cropper}, {Crosta}, {Crowley}, {Dafonte}, {Dapergolas}, {David}, {de Laverny}, {De Luise}, {De March}, {De Ridder}, {de Souza}, {de Torres}, {del Peloso}, {del Pozo}, {Delbo}, {Delgado}, {Delisle}, {Demouchy}, {Dharmawardena}, {Diakite}, {Diener}, {Distefano}, {Dolding}, {Enke}, {Fabre}, {Fabrizio}, {Faigler}, {Fedorets}, {Fernique}, {Figueras}, {Fournier}, {Fouron}, {Fragkoudi}, {Gai}, {Garcia-Gutierrez}, {Garcia-Reinaldos}, {Garc{\'\i}a-Torres}, {Garofalo}, {Gavel}, {Gavras}, {Gerlach}, {Geyer}, {Giacobbe}, {Gilmore}, {Girona}, {Giuffrida}, {Gomel}, {Gomez}, {Gonz{\'a}lez-N{\'u}{\~n}ez},
  {Gonz{\'a}lez-Santamar{\'\i}a}, {Gonz{\'a}lez-Vidal}, {Granvik}, {Guillout}, {Guiraud}, {Guti{\'e}rrez-S{\'a}nchez}, {Guy}, {Hatzidimitriou}, {Hauser}, {Haywood}, {Helmer}, {Helmi}, {Sarmiento}, {Hidalgo}, {H{\l}adczuk}, {Hobbs}, {Holland}, {Huckle}, {Jardine}, {Jasniewicz}, {Jean-Antoine Piccolo}, {Jim{\'e}nez-Arranz}, {Juaristi Campillo}, {Julbe}, {Karbevska}, {Kervella}, {Khanna}, {Kordopatis}, {Korn}, {K{\'o}sp{\'a}l}, {Kostrzewa-Rutkowska}, {Kruszy{\'n}ska}, {Kun}, {Laizeau}, {Lambert}, {Lanza}, {Lasne}, {Le Campion}, {Lebreton}, {Lebzelter}, {Leccia}, {Leclerc}, {Lecoeur-Taibi}, {Liao}, {Licata}, {Lindstr{\'o}m}, {Lister}, {Livanou}, {Lobel}, {Lorca}, {Loup}, {Madrero Pardo}, {Magdaleno Romeo}, {Managau}, {Mann}, {Marchant}, {Marconi}, {Marcos}, {Marcos Santos}, {Mar{\'\i}n Pina}, {Marinoni}, {Marocco}, {Marshall}, {Martin Polo}, {Mart{\'\i}n-Fleitas}, {Marton}, {Mary}, {Masip}, {Mastrobuono-Battisti}, {Mazeh}, {McMillan}, {Messina}, {Michalik}, {Millar}, {Mints}, {Molina}, {Molinaro}, {Moln{\'a}r},
  {Monari}, {Mongui{\'o}}, {Montero}, {Mor}, {Mora}, {Morbidelli}, {Morel}, {Morris}, {Muraveva}, {Murphy}, {Musella}, {Nagy}, {Noval}, {Oca{\~n}a}, {Ogden}, {Ordenovic}, {Osinde}, {Pagani}, {Pagano}, {Palicio}, {Pallas-Quintela}, {Panahi}, {Payne-Wardenaar}, {Pe{\~n}alosa Esteller}, {Penttil{\"a}}, {Pichon}, {Piersimoni}, {Pineau}, {Plachy}, {Plum}, {Poggio}, {Pr{\v{s}}a}, {Pulone}, {Racero}, {Ragaini}, {Rainer}, {Raiteri}, {Ramos}, {Ramos-Lerate}, {Re Fiorentin}, {Regibo}, {Richards}, {Rios Diaz}, {Ripepi}, {Riva}, {Rix}, {Rixon}, {Robichon}, {Robin}, {Robin}, {Roelens}, {Rogues}, {Rohrbasser}, {Romero-G{\'o}mez}, {Rowell}, {Royer}, {Rybicki}, {Sadowski}, {S{\'a}ez N{\'u}{\~n}ez}, {Sagrist{\`a} Sell{\'e}s}, {Sahlmann}, {Salguero}, {Samaras}, {Sanchez Gimenez}, {Sarasso}, {Schultheis}, {Sciacca}, {Segol}, {Segovia}, {S{\'e}gransan}, {Semeux}, {Shahaf}, {Siddiqui}, {Siebert}, {Siltala}, {Silvelo}, {Slezak}, {Slezak}, {Smart}, {Snaith}, {Solano}, {Solitro}, {Souami}, {Souchay}, {Spagna}, {Spina}, {Spoto},
  {Steele}, {Steidelm{\"u}ller}, {Stephenson}, {S{\"u}veges}, {Surdej}, {Szabados}, {Szegedi-Elek}, {Taris}, {Taylor}, {Teixeira}, {Tolomei}, {Tonello}, {Torra}, {Torra}, {Torralba Elipe}, {Trabucchi}, {Tsounis}, {Turon}, {Ulla}, {Unger}, {Vaillant}, {van Dillen}, {van Reeven}, {Vanel}, {Vecchiato}, {Viala}, {Vicente}, {Voutsinas}, {Wevers}, {Wyrzykowski}, {Yoldas}, {Yvard}, {Zhao}, {Zorec}, {Zucker}, \& {Zwitter}}]{GaiaCollaboration2023b}
{Gaia Collaboration}, {Montegriffo}, P., {Bellazzini}, M., {et~al.} 2023, \aap, 674, A33, \dodoi{10.1051/0004-6361/202243709}

\bibitem[{{Gilmore} {et~al.}(2022){Gilmore}, {Randich}, {Worley}, {Hourihane}, {Gonneau}, {Sacco}, {Lewis}, {Magrini}, {Fran{\c{c}}ois}, {Jeffries}, {Koposov}, {Bragaglia}, {Alfaro}, {Allende Prieto}, {Blomme}, {Korn}, {Lanzafame}, {Pancino}, {Recio-Blanco}, {Smiljanic}, {Van Eck}, {Zwitter}, {Bensby}, {Flaccomio}, {Irwin}, {Franciosini}, {Morbidelli}, {Damiani}, {Bonito}, {Friel}, {Vink}, {Prisinzano}, {Abbas}, {Hatzidimitriou}, {Held}, {Jordi}, {Paunzen}, {Spagna}, {Jackson}, {Ma{\'\i}z Apell{\'a}niz}, {Asplund}, {Bonifacio}, {Feltzing}, {Binney}, {Drew}, {Ferguson}, {Micela}, {Negueruela}, {Prusti}, {Rix}, {Vallenari}, {Bergemann}, {Casey}, {de Laverny}, {Frasca}, {Hill}, {Lind}, {Sbordone}, {Sousa}, {Adibekyan}, {Caffau}, {Daflon}, {Feuillet}, {Gebran}, {Gonzalez Hernandez}, {Guiglion}, {Herrero}, {Lobel}, {Merle}, {Mikolaitis}, {Montes}, {Morel}, {Ruchti}, {Soubiran}, {Tabernero}, {Tautvai{\v{s}}ien{\.{e}}}, {Traven}, {Valentini}, {Van der Swaelmen}, {Villanova}, {Viscasillas V{\'a}zquez}, {Bayo},
  {Biazzo}, {Carraro}, {Edvardsson}, {Heiter}, {Jofr{\'e}}, {Marconi}, {Martayan}, {Masseron}, {Monaco}, {Walton}, {Zaggia}, {Aguirre B{\o}rsen-Koch}, {Alves}, {Balaguer-Nunez}, {Barklem}, {Barrado}, {Bellazzini}, {Berlanas}, {Binks}, {Bressan}, {Capuzzo-Dolcetta}, {Casagrande}, {Casamiquela}, {Collins}, {D'Orazi}, {Dantas}, {Debattista}, {Delgado-Mena}, {Di Marcantonio}, {Drazdauskas}, {Evans}, {Famaey}, {Franchini}, {Fr{\'e}mat}, {Fu}, {Geisler}, {Gerhard}, {Gonz{\'a}lez Solares}, {Grebel}, {Guti{\'e}rrez Albarr{\'a}n}, {Jim{\'e}nez-Esteban}, {J{\"o}nsson}, {Khachaturyants}, {Kordopatis}, {Kos}, {Lagarde}, {Ludwig}, {Mahy}, {Mapelli}, {Marfil}, {Martell}, {Messina}, {Miglio}, {Minchev}, {Moitinho}, {Montalban}, {Monteiro}, {Morossi}, {Mowlavi}, {Mucciarelli}, {Murphy}, {Nardetto}, {Ortolani}, {Paletou}, {Palou{\v{s}}}, {Pickering}, {Quirrenbach}, {Re Fiorentin}, {Read}, {Romano}, {Ryde}, {Sanna}, {Santos}, {Seabroke}, {Spina}, {Steinmetz}, {Stonkut{\'e}}, {Sutorius}, {Th{\'e}venin}, {Tosi}, {Tsantaki},
  {Wright}, {Wyse}, {Zoccali}, {Zorec}, \& {Zucker}}]{gilmore2022}
{Gilmore}, G., {Randich}, S., {Worley}, C.~C., {et~al.} 2022, \aap, 666, A120, \dodoi{10.1051/0004-6361/202243134}

\bibitem[{{GRAVITY Collaboration} {et~al.}(2020){GRAVITY Collaboration}, {Abuter}, {Amorim}, {Baub{\"o}ck}, {Berger}, {Bonnet}, {Brandner}, {Cardoso}, {Cl{\'e}net}, {de Zeeuw}, {Dexter}, {Eckart}, {Eisenhauer}, {F{\"o}rster Schreiber}, {Garcia}, {Gao}, {Gendron}, {Genzel}, {Gillessen}, {Habibi}, {Haubois}, {Henning}, {Hippler}, {Horrobin}, {Jim{\'e}nez-Rosales}, {Jochum}, {Jocou}, {Kaufer}, {Kervella}, {Lacour}, {Lapeyr{\`e}re}, {Le Bouquin}, {L{\'e}na}, {Nowak}, {Ott}, {Paumard}, {Perraut}, {Perrin}, {Pfuhl}, {Rodr{\'\i}guez-Coira}, {Shangguan}, {Scheithauer}, {Stadler}, {Straub}, {Straubmeier}, {Sturm}, {Tacconi}, {Vincent}, {von Fellenberg}, {Waisberg}, {Widmann}, {Wieprecht}, {Wiezorrek}, {Woillez}, {Yazici}, \& {Zins}}]{GRAVITYCollaboration2020}
{GRAVITY Collaboration}, {Abuter}, R., {Amorim}, A., {et~al.} 2020, \aap, 636, L5, \dodoi{10.1051/0004-6361/202037813}

\bibitem[{{Harris}(2010)}]{Harris2010}
{Harris}, W.~E. 2010, arXiv e-prints, arXiv:1012.3224, \dodoi{10.48550/arXiv.1012.3224}

\bibitem[{{Haywood} {et~al.}(2018){Haywood}, {Di Matteo}, {Lehnert}, {Snaith}, {Khoperskov}, \& {G{\'o}mez}}]{Haywood2018}
{Haywood}, M., {Di Matteo}, P., {Lehnert}, M.~D., {et~al.} 2018, \apj, 863, 113, \dodoi{10.3847/1538-4357/aad235}

\bibitem[{{Helmi} {et~al.}(2018){Helmi}, {Babusiaux}, {Koppelman}, {Massari}, {Veljanoski}, \& {Brown}}]{Helmi2018}
{Helmi}, A., {Babusiaux}, C., {Koppelman}, H.~H., {et~al.} 2018, \nat, 563, 85, \dodoi{10.1038/s41586-018-0625-x}

\bibitem[{{Hirai} {et~al.}(2022){Hirai}, {Beers}, {Chiba}, {Aoki}, {Shank}, {Saitoh}, {Okamoto}, \& {Makino}}]{Hirai2022}
{Hirai}, Y., {Beers}, T.~C., {Chiba}, M., {et~al.} 2022, \mnras, 517, 4856, \dodoi{10.1093/mnras/stac2489}

\bibitem[{{Huang} {et~al.}(2015){Huang}, {Liu}, {Yuan}, {Xiang}, {Chen}, \& {Zhang}}]{Huang2015b}
{Huang}, Y., {Liu}, X.~W., {Yuan}, H.~B., {et~al.} 2015, \mnras, 454, 2863, \dodoi{10.1093/mnras/stv1991}

\bibitem[{{Huang} {et~al.}(2021){Huang}, {Yuan}, {Li}, {Wolf}, {Onken}, {Beers}, {Casagrande}, {Mackey}, {Da Costa}, {Bland-Hawthorn}, {Stello}, {Nordlander}, {Ting}, {Buder}, {Sharma}, \& {Liu}}]{Huang2021a}
{Huang}, Y., {Yuan}, H., {Li}, C., {et~al.} 2021, \apj, 907, 68, \dodoi{10.3847/1538-4357/abca37}

\bibitem[{{Huang} {et~al.}(2022){Huang}, {Beers}, {Wolf}, {Lee}, {Onken}, {Yuan}, {Shank}, {Zhang}, {Wang}, {Shi}, \& {Fan}}]{Huang2022}
{Huang}, Y., {Beers}, T.~C., {Wolf}, C., {et~al.} 2022, \apj, 925, 164, \dodoi{10.3847/1538-4357/ac21cb}

\bibitem[{{Huang} {et~al.}(2023){Huang}, {Beers}, {Yuan}, {Tan}, {Wang}, {Zheng}, {Li}, {Lee}, {Li}, {Zhao}, {Xue}, {Liu}, {Zhang}, {Sun}, {Li}, {Gu}, {Wolf}, {Onken}, {Liu}, {Fan}, \& {Zhao}}]{Huang2023}
{Huang}, Y., {Beers}, T.~C., {Yuan}, H., {et~al.} 2023, \apj, 957, 65, \dodoi{10.3847/1538-4357/ace628}

\bibitem[{{Keller} {et~al.}(2007){Keller}, {Schmidt}, {Bessell}, {Conroy}, {Francis}, {Granlund}, {Kowald}, {Oates}, {Martin-Jones}, {Preston}, {Tisserand}, {Vaccarella}, \& {Waterson}}]{Keller2007}
{Keller}, S.~C., {Schmidt}, B.~P., {Bessell}, M.~S., {et~al.} 2007, \pasa, 24, 1, \dodoi{10.1071/AS07001}

\bibitem[{{Kim} {et~al.}(2021){Kim}, {Lee}, {Beers}, \& {Koo}}]{Kim2021}
{Kim}, Y.~K., {Lee}, Y.~S., {Beers}, T.~C., \& {Koo}, J.-R. 2021, \apjl, 911, L21, \dodoi{10.3847/2041-8213/abf35e}

\bibitem[{{Koppelman} {et~al.}(2021){Koppelman}, {Hagen}, \& {Helmi}}]{Koppelman2021}
{Koppelman}, H.~H., {Hagen}, J. H.~J., \& {Helmi}, A. 2021, \aap, 647, A37, \dodoi{10.1051/0004-6361/202039390}

\bibitem[{{Lee} {et~al.}(2023){Lee}, {Lee}, {Kim}, {Beers}, \& {An}}]{Lee2023}
{Lee}, A., {Lee}, Y.~S., {Kim}, Y.~K., {Beers}, T.~C., \& {An}, D. 2023, \apj, 945, 56, \dodoi{10.3847/1538-4357/acb6f5}

\bibitem[{{Li} {et~al.}(2022{\natexlab{a}}){Li}, {Aoki}, {Matsuno}, {Xing}, {Suda}, {Tominaga}, {Chen}, {Honda}, {Ishigaki}, {Shi}, {Zhao}, \& {Zhao}}]{Li2022b}
{Li}, H., {Aoki}, W., {Matsuno}, T., {et~al.} 2022{\natexlab{a}}, \apj, 931, 147, \dodoi{10.3847/1538-4357/ac6514}

\bibitem[{{Li} {et~al.}(2022{\natexlab{b}}){Li}, {Ji}, {Pace}, {Erkal}, {Koposov}, {Shipp}, {Da Costa}, {Cullinane}, {Kuehn}, {Lewis}, {Mackey}, {Simpson}, {Zucker}, {Ferguson}, {Martell}, {Bland-Hawthorn}, {Balbinot}, {Tavangar}, {Drlica-Wagner}, {De Silva}, \& {Simon}}]{Li2022a}
{Li}, T.~S., {Ji}, A.~P., {Pace}, A.~B., {et~al.} 2022{\natexlab{b}}, \apj, 928, 30, \dodoi{10.3847/1538-4357/ac46d3}

\bibitem[{{Limberg} {et~al.}(2021){Limberg}, {Santucci}, {Rossi}, {Shank}, {Placco}, {Beers}, {Schlaufman}, {Casey}, {Perottoni}, \& {Lee}}]{Limberg2021}
{Limberg}, G., {Santucci}, R.~M., {Rossi}, S., {et~al.} 2021, \apj, 913, 11, \dodoi{10.3847/1538-4357/abeefe}

\bibitem[{{Luo} {et~al.}(2015){Luo}, {Zhao}, {Zhao}, {Deng}, {Liu}, {Jing}, {Wang}, {Zhang}, {Shi}, {Cui}, {Chu}, {Li}, {Bai}, {Wu}, {Cai}, {Cao}, {Cao}, {Carlin}, {Chen}, {Chen}, {Chen}, {Chen}, {Chen}, {Chen}, {Chen}, {Christlieb}, {Chu}, {Cui}, {Dong}, {Du}, {Fan}, {Feng}, {Fu}, {Gao}, {Gong}, {Gu}, {Guo}, {Han}, {He}, {Hou}, {Hou}, {Hou}, {Hu}, {Hu}, {Hu}, {Huo}, {Jia}, {Jiang}, {Jiang}, {Jiang}, {Jin}, {Kong}, {Kong}, {Lei}, {Li}, {Li}, {Li}, {Li}, {Li}, {Li}, {Li}, {Li}, {Li}, {Li}, {Li}, {Li}, {Liang}, {Lin}, {Liu}, {Liu}, {Liu}, {Liu}, {Lu}, {Luo}, {Mao}, {Newberg}, {Ni}, {Qi}, {Qi}, {Shen}, {Shi}, {Song}, {Song}, {Su}, {Su}, {Tang}, {Tao}, {Tian}, {Wang}, {Wang}, {Wang}, {Wang}, {Wang}, {Wang}, {Wang}, {Wang}, {Wang}, {Wang}, {Wang}, {Wang}, {Wang}, {Wang}, {Wang}, {Wang}, {Wang}, {Wang}, {Wang}, {Wang}, {Wei}, {Wei}, {Wu}, {Wu}, {Wu}, {Wu}, {Xing}, {Xu}, {Xu}, {Xu}, {Yan}, {Yang}, {Yang}, {Yang}, {Yang}, {Yao}, {Yu}, {Yuan}, {Yuan}, {Yuan}, {Yuan}, {Zhai}, {Zhang}, {Zhang}, {Zhang}, {Zhang},
  {Zhang}, {Zhang}, {Zhang}, {Zhang}, {Zhao}, {Zhou}, {Zhou}, {Zhu}, {Zhu}, {Zou}, \& {Zuo}}]{Luo2015}
{Luo}, A.~L., {Zhao}, Y.-H., {Zhao}, G., {et~al.} 2015, Research in Astronomy and Astrophysics, 15, 1095, \dodoi{10.1088/1674-4527/15/8/002}

\bibitem[{{Majewski} {et~al.}(2017){Majewski}, {Schiavon}, {Frinchaboy}, {Allende Prieto}, {Barkhouser}, {Bizyaev}, {Blank}, {Brunner}, {Burton}, {Carrera}, {Chojnowski}, {Cunha}, {Epstein}, {Fitzgerald}, {Garc{\'\i}a P{\'e}rez}, {Hearty}, {Henderson}, {Holtzman}, {Johnson}, {Lam}, {Lawler}, {Maseman}, {M{\'e}sz{\'a}ros}, {Nelson}, {Nguyen}, {Nidever}, {Pinsonneault}, {Shetrone}, {Smee}, {Smith}, {Stolberg}, {Skrutskie}, {Walker}, {Wilson}, {Zasowski}, {Anders}, {Basu}, {Beland}, {Blanton}, {Bovy}, {Brownstein}, {Carlberg}, {Chaplin}, {Chiappini}, {Eisenstein}, {Elsworth}, {Feuillet}, {Fleming}, {Galbraith-Frew}, {Garc{\'\i}a}, {Garc{\'\i}a-Hern{\'a}ndez}, {Gillespie}, {Girardi}, {Gunn}, {Hasselquist}, {Hayden}, {Hekker}, {Ivans}, {Kinemuchi}, {Klaene}, {Mahadevan}, {Mathur}, {Mosser}, {Muna}, {Munn}, {Nichol}, {O'Connell}, {Parejko}, {Robin}, {Rocha-Pinto}, {Schultheis}, {Serenelli}, {Shane}, {Silva Aguirre}, {Sobeck}, {Thompson}, {Troup}, {Weinberg}, \& {Zamora}}]{Majewski2017}
{Majewski}, S.~R., {Schiavon}, R.~P., {Frinchaboy}, P.~M., {et~al.} 2017, \aj, 154, 94, \dodoi{10.3847/1538-3881/aa784d}

\bibitem[{{Mardini} {et~al.}(2024){Mardini}, {Frebel}, \& {Chiti}}]{Mardini2024}
{Mardini}, M.~K., {Frebel}, A., \& {Chiti}, A. 2024, \mnras, 529, L60, \dodoi{10.1093/mnrasl/slad197}

\bibitem[{{Mardini} {et~al.}(2022{\natexlab{a}}){Mardini}, {Frebel}, {Chiti}, {Meiron}, {Brauer}, \& {Ou}}]{Mardini2022a}
{Mardini}, M.~K., {Frebel}, A., {Chiti}, A., {et~al.} 2022{\natexlab{a}}, \apj, 936, 78, \dodoi{10.3847/1538-4357/ac8102}

\bibitem[{{Mardini} {et~al.}(2022{\natexlab{b}}){Mardini}, {Frebel}, {Ezzeddine}, {Chiti}, {Meiron}, {Ji}, {Placco}, {Roederer}, \& {Mel{\'e}ndez}}]{Mardini2022b}
{Mardini}, M.~K., {Frebel}, A., {Ezzeddine}, R., {et~al.} 2022{\natexlab{b}}, \mnras, 517, 3993, \dodoi{10.1093/mnras/stac2783}

\bibitem[{{Marigo} {et~al.}(2017){Marigo}, {Girardi}, {Bressan}, {Rosenfield}, {Aringer}, {Chen}, {Dussin}, {Nanni}, {Pastorelli}, {Rodrigues}, {Trabucchi}, {Bladh}, {Dalcanton}, {Groenewegen}, {Montalb{\'a}n}, \& {Wood}}]{Marigo2017}
{Marigo}, P., {Girardi}, L., {Bressan}, A., {et~al.} 2017, \apj, 835, 77, \dodoi{10.3847/1538-4357/835/1/77}

\bibitem[{{McMillan}(2017)}]{McMillan2017}
{McMillan}, P.~J. 2017, \mnras, 465, 76, \dodoi{10.1093/mnras/stw2759}

\bibitem[{{Mendes de Oliveira} {et~al.}(2019){Mendes de Oliveira}, {Ribeiro}, {Schoenell}, {Kanaan}, {Overzier}, {Molino}, {Sampedro}, {Coelho}, {Barbosa}, {Cortesi}, {Costa-Duarte}, {Herpich}, {Hernandez-Jimenez}, {Placco}, {Xavier}, {Abramo}, {Saito}, {Chies-Santos}, {Ederoclite}, {Lopes de Oliveira}, {Gon{\c{c}}alves}, {Akras}, {Almeida}, {Almeida-Fernandes}, {Beers}, {Bonatto}, {Bonoli}, {Cypriano}, {Vinicius-Lima}, {de Souza}, {Fabiano de Souza}, {Ferrari}, {Gon{\c{c}}alves}, {Gonzalez}, {Guti{\'e}rrez-Soto}, {Hartmann}, {Jaffe}, {Kerber}, {Lima-Dias}, {Lopes}, {Menendez-Delmestre}, {Nakazono}, {Novais}, {Ortega-Minakata}, {Pereira}, {Perottoni}, {Queiroz}, {Reis}, {Santos}, {Santos-Silva}, {Santucci}, {Barbosa}, {Siffert}, {Sodr{\'e}}, {Torres-Flores}, {Westera}, {Whitten}, {Alcaniz}, {Alonso-Garc{\'\i}a}, {Alencar}, {Alvarez-Candal}, {Amram}, {Azanha}, {Barb{\'a}}, {Bernardinelli}, {Borges Fernandes}, {Branco}, {Brito-Silva}, {Buzzo}, {Caffer}, {Campillay}, {Cano}, {Carvano}, {Castejon}, {Cid
  Fernandes}, {Dantas}, {Daflon}, {Damke}, {de la Reza}, {de Melo de Azevedo}, {De Paula}, {Diem}, {Donnerstein}, {Dors}, {Dupke}, {Eikenberry}, {Escudero}, {Faifer}, {Far{\'\i}as}, {Fernandes}, {Fernandes}, {Fontes}, {Galarza}, {Hirata}, {Katena}, {Gregorio-Hetem}, {Hern{\'a}ndez-Fern{\'a}ndez}, {Izzo}, {Jaque Arancibia}, {Jatenco-Pereira}, {Jim{\'e}nez-Teja}, {Kann}, {Krabbe}, {Labayru}, {Lazzaro}, {Lima Neto}, {Lopes}, {Magalh{\~a}es}, {Makler}, {de Menezes}, {Miralda-Escud{\'e}}, {Monteiro-Oliveira}, {Montero-Dorta}, {Mu{\~n}oz-Elgueta}, {Nemmen}, {Nilo Castell{\'o}n}, {Oliveira}, {Ort{\'\i}z}, {Pattaro}, {Pereira}, {Quint}, {Riguccini}, {Rocha Pinto}, {Rodrigues}, {Roig}, {Rossi}, {Saha}, {Santos}, {Schnorr M{\"u}ller}, {Sesto}, {Silva}, {Smith Castelli}, {Teixeira}, {Telles}, {Thom de Souza}, {Th{\"o}ne}, {Trevisan}, {de Ugarte Postigo}, {Urrutia-Viscarra}, {Veiga}, {Vika}, {Vitorelli}, {Werle}, {Werner}, \& {Zaritsky}}]{MendesdeOliveira2019}
{Mendes de Oliveira}, C., {Ribeiro}, T., {Schoenell}, W., {et~al.} 2019, \mnras, 489, 241, \dodoi{10.1093/mnras/stz1985}

\bibitem[{{Norris} {et~al.}(1985){Norris}, {Bessell}, \& {Pickles}}]{Norris1985}
{Norris}, J., {Bessell}, M.~S., \& {Pickles}, A.~J. 1985, \apjs, 58, 463, \dodoi{10.1086/191049}

\bibitem[{{Onken} {et~al.}(2019){Onken}, {Wolf}, {Bessell}, {Chang}, {Da Costa}, {Luvaul}, {Mackey}, {Schmidt}, \& {Shao}}]{Onken2019}
{Onken}, C.~A., {Wolf}, C., {Bessell}, M.~S., {et~al.} 2019, \pasa, 36, e033, \dodoi{10.1017/pasa.2019.27}

\bibitem[{{Placco} {et~al.}(2022){Placco}, {Almeida-Fernandes}, {Arentsen}, {Lee}, {Schoenell}, {Ribeiro}, \& {Kanaan}}]{Placco2022}
{Placco}, V.~M., {Almeida-Fernandes}, F., {Arentsen}, A., {et~al.} 2022, \apjs, 262, 8, \dodoi{10.3847/1538-4365/ac7ab0}

\bibitem[{{Placco} {et~al.}(2014){Placco}, {Frebel}, {Beers}, \& {Stancliffe}}]{Placco2014}
{Placco}, V.~M., {Frebel}, A., {Beers}, T.~C., \& {Stancliffe}, R.~J. 2014, \apj, 797, 21, \dodoi{10.1088/0004-637X/797/1/21}

\bibitem[{{Reid} \& {Brunthaler}(2020)}]{Reid2020}
{Reid}, M.~J., \& {Brunthaler}, A. 2020, \apj, 892, 39, \dodoi{10.3847/1538-4357/ab76cd}

\bibitem[{{Rockosi} {et~al.}(2022){Rockosi}, {Lee}, {Morrison}, {Yanny}, {Johnson}, {Lucatello}, {Sobeck}, {Beers}, {Allende Prieto}, {An}, {Bizyaev}, {Blanton}, {Casagrande}, {Eisenstein}, {Gould}, {Gunn}, {Harding}, {Ivans}, {Jacobson}, {Janesh}, {Knapp}, {Kollmeier}, {L{\'e}pine}, {L{\'o}pez-Corredoira}, {Ma}, {Newberg}, {Pan}, {Prchlik}, {Sayers}, {Schlesinger}, {Simmerer}, \& {Weinberg}}]{Rockosi2022}
{Rockosi}, C.~M., {Lee}, Y.~S., {Morrison}, H.~L., {et~al.} 2022, \apjs, 259, 60, \dodoi{10.3847/1538-4365/ac5323}

\bibitem[{{Rybizki} {et~al.}(2018){Rybizki}, {Demleitner}, {Fouesneau}, {Bailer-Jones}, {Rix}, \& {Andrae}}]{Rybizki2018}
{Rybizki}, J., {Demleitner}, M., {Fouesneau}, M., {et~al.} 2018, \pasp, 130, 074101, \dodoi{10.1088/1538-3873/aabd70}

\bibitem[{{Santistevan} {et~al.}(2021){Santistevan}, {Wetzel}, {Sanderson}, {El-Badry}, {Samuel}, \& {Faucher-Gigu{\`e}re}}]{Santistevan2021}
{Santistevan}, I.~B., {Wetzel}, A., {Sanderson}, R.~E., {et~al.} 2021, \mnras, 505, 921, \dodoi{10.1093/mnras/stab1345}

\bibitem[{{Schlaufman} {et~al.}(2018){Schlaufman}, {Thompson}, \& {Casey}}]{Schlaufman2018}
{Schlaufman}, K.~C., {Thompson}, I.~B., \& {Casey}, A.~R. 2018, \apj, 867, 98, \dodoi{10.3847/1538-4357/aadd97}

\bibitem[{{Schlegel} {et~al.}(1998){Schlegel}, {Finkbeiner}, \& {Davis}}]{Schlegel1998}
{Schlegel}, D.~J., {Finkbeiner}, D.~P., \& {Davis}, M. 1998, \apj, 500, 525, \dodoi{10.1086/305772}

\bibitem[{{Sch{\"o}nrich} {et~al.}(2010){Sch{\"o}nrich}, {Binney}, \& {Dehnen}}]{Schonrich2010}
{Sch{\"o}nrich}, R., {Binney}, J., \& {Dehnen}, W. 2010, \mnras, 403, 1829, \dodoi{10.1111/j.1365-2966.2010.16253.x}

\bibitem[{{Schuster} {et~al.}(2012){Schuster}, {Moreno}, {Nissen}, \& {Pichardo}}]{Schuster2012}
{Schuster}, W.~J., {Moreno}, E., {Nissen}, P.~E., \& {Pichardo}, B. 2012, \aap, 538, A21, \dodoi{10.1051/0004-6361/201118035}

\bibitem[{{Sestito} {et~al.}(2019){Sestito}, {Martin}, \& {Starkenburg}}]{Sestito2019}
{Sestito}, F., {Martin}, N., \& {Starkenburg}, E. 2019, in The Gaia Universe, 47, \dodoi{10.5281/zenodo.3236051}

\bibitem[{{Sestito} {et~al.}(2020){Sestito}, {Martin}, {Starkenburg}, {Arentsen}, {Ibata}, {Longeard}, {Kielty}, {Youakim}, {Venn}, {Aguado}, {Carlberg}, {Gonz{\'a}lez Hern{\'a}ndez}, {Hill}, {Jablonka}, {Kordopatis}, {Malhan}, {Navarro}, {S{\'a}nchez-Janssen}, {Thomas}, {Tolstoy}, {Wilson}, {Palicio}, {Bialek}, {Garcia-Dias}, {Lucchesi}, {North}, {Osorio}, {Patrick}, \& {Peralta de Arriba}}]{Sestito2020}
{Sestito}, F., {Martin}, N.~F., {Starkenburg}, E., {et~al.} 2020, \mnras, 497, L7, \dodoi{10.1093/mnrasl/slaa022}

\bibitem[{{Sestito} {et~al.}(2021){Sestito}, {Buck}, {Starkenburg}, {Martin}, {Navarro}, {Venn}, {Obreja}, {Jablonka}, \& {Macci{\`o}}}]{Sestito2021}
{Sestito}, F., {Buck}, T., {Starkenburg}, E., {et~al.} 2021, \mnras, 500, 3750, \dodoi{10.1093/mnras/staa3479}

\bibitem[{{Shank} {et~al.}(2022){Shank}, {Komater}, {Beers}, {Placco}, \& {Huang}}]{Shank2022b}
{Shank}, D., {Komater}, D., {Beers}, T.~C., {Placco}, V.~M., \& {Huang}, Y. 2022, \apjs, 261, 19, \dodoi{10.3847/1538-4365/ac680c}

\bibitem[{{Sotillo-Ramos} {et~al.}(2023){Sotillo-Ramos}, {Bergemann}, {Friske}, \& {Pillepich}}]{Sotillo-Ramos2023}
{Sotillo-Ramos}, D., {Bergemann}, M., {Friske}, J. K.~S., \& {Pillepich}, A. 2023, \mnras, 525, L105, \dodoi{10.1093/mnrasl/slad103}

\bibitem[{{Starkenburg} {et~al.}(2017){Starkenburg}, {Martin}, {Youakim}, {Aguado}, {Allende Prieto}, {Arentsen}, {Bernard}, {Bonifacio}, {Caffau}, {Carlberg}, {C{\^o}t{\'e}}, {Fouesneau}, {Fran{\c{c}}ois}, {Franke}, {Gonz{\'a}lez Hern{\'a}ndez}, {Gwyn}, {Hill}, {Ibata}, {Jablonka}, {Longeard}, {McConnachie}, {Navarro}, {S{\'a}nchez-Janssen}, {Tolstoy}, \& {Venn}}]{Starkenburg2017}
{Starkenburg}, E., {Martin}, N., {Youakim}, K., {et~al.} 2017, \mnras, 471, 2587, \dodoi{10.1093/mnras/stx1068}

\bibitem[{{Steinmetz} {et~al.}(2006){Steinmetz}, {Zwitter}, {Siebert}, {Watson}, {Freeman}, {Munari}, {Campbell}, {Williams}, {Seabroke}, {Wyse}, {Parker}, {Bienaym{\'e}}, {Roeser}, {Gibson}, {Gilmore}, {Grebel}, {Helmi}, {Navarro}, {Burton}, {Cass}, {Dawe}, {Fiegert}, {Hartley}, {Russell}, {Saunders}, {Enke}, {Bailin}, {Binney}, {Bland-Hawthorn}, {Boeche}, {Dehnen}, {Eisenstein}, {Evans}, {Fiorucci}, {Fulbright}, {Gerhard}, {Jauregi}, {Kelz}, {Mijovi{\'c}}, {Minchev}, {Parmentier}, {Pe{\~n}arrubia}, {Quillen}, {Read}, {Ruchti}, {Scholz}, {Siviero}, {Smith}, {Sordo}, {Veltz}, {Vidrih}, {von Berlepsch}, {Boyle}, \& {Schilbach}}]{Steinmetz2006}
{Steinmetz}, M., {Zwitter}, T., {Siebert}, A., {et~al.} 2006, \aj, 132, 1645, \dodoi{10.1086/506564}

\bibitem[{{Vasiliev}(2019)}]{Vasiliev2019}
{Vasiliev}, E. 2019, \mnras, 482, 1525, \dodoi{10.1093/mnras/sty2672}

\bibitem[{{Venn} {et~al.}(2020){Venn}, {Kielty}, {Sestito}, {Starkenburg}, {Martin}, {Aguado}, {Arentsen}, {Bonifacio}, {Caffau}, {Hill}, {Jablonka}, {Lardo}, {Mashonkina}, {Navarro}, {Sneden}, {Thomas}, {Youakim}, {Gonz{\'a}lez-Hern{\'a}ndez}, {S{\'a}nchez Janssen}, {Carlberg}, \& {Malhan}}]{Venn2020}
{Venn}, K.~A., {Kielty}, C.~L., {Sestito}, F., {et~al.} 2020, \mnras, 492, 3241, \dodoi{10.1093/mnras/stz3546}

\bibitem[{{Viswanathan} {et~al.}(2024){Viswanathan}, {Starkenburg}, {Matsuno}, {Venn}, {Martin}, {Longeard}, {Ardern-Arentsen}, {Carlberg}, {Fabbro}, {Kordopatis}, {Montelius}, {Sestito}, \& {Yuan}}]{Viswanathan2024}
{Viswanathan}, A., {Starkenburg}, E., {Matsuno}, T., {et~al.} 2024, \aap, 683, L11, \dodoi{10.1051/0004-6361/202347944}

\bibitem[{{Xu} {et~al.}(2022){Xu}, {Yuan}, {Niu}, {Yang}, {Beers}, \& {Huang}}]{Xu2022}
{Xu}, S., {Yuan}, H., {Niu}, Z., {et~al.} 2022, \apjs, 258, 44, \dodoi{10.3847/1538-4365/ac3df6}

\bibitem[{{Yanny} {et~al.}(2009){Yanny}, {Rockosi}, {Newberg}, {Knapp}, {Adelman-McCarthy}, {Alcorn}, {Allam}, {Allende Prieto}, {An}, {Anderson}, {Anderson}, {Bailer-Jones}, {Bastian}, {Beers}, {Bell}, {Belokurov}, {Bizyaev}, {Blythe}, {Bochanski}, {Boroski}, {Brinchmann}, {Brinkmann}, {Brewington}, {Carey}, {Cudworth}, {Evans}, {Evans}, {Gates}, {G{\"a}nsicke}, {Gillespie}, {Gilmore}, {Nebot Gomez-Moran}, {Grebel}, {Greenwell}, {Gunn}, {Jordan}, {Jordan}, {Harding}, {Harris}, {Hendry}, {Holder}, {Ivans}, {Ivezi{\v{c}}}, {Jester}, {Johnson}, {Kent}, {Kleinman}, {Kniazev}, {Krzesinski}, {Kron}, {Kuropatkin}, {Lebedeva}, {Lee}, {French Leger}, {L{\'e}pine}, {Levine}, {Lin}, {Long}, {Loomis}, {Lupton}, {Malanushenko}, {Malanushenko}, {Margon}, {Martinez-Delgado}, {McGehee}, {Monet}, {Morrison}, {Munn}, {Neilsen}, {Nitta}, {Norris}, {Oravetz}, {Owen}, {Padmanabhan}, {Pan}, {Peterson}, {Pier}, {Platson}, {Re Fiorentin}, {Richards}, {Rix}, {Schlegel}, {Schneider}, {Schreiber}, {Schwope}, {Sibley}, {Simmons},
  {Snedden}, {Allyn Smith}, {Stark}, {Stauffer}, {Steinmetz}, {Stoughton}, {SubbaRao}, {Szalay}, {Szkody}, {Thakar}, {Sivarani}, {Tucker}, {Uomoto}, {Vanden Berk}, {Vidrih}, {Wadadekar}, {Watters}, {Wilhelm}, {Wyse}, {Yarger}, \& {Zucker}}]{Yanny2009}
{Yanny}, B., {Rockosi}, C., {Newberg}, H.~J., {et~al.} 2009, \aj, 137, 4377, \dodoi{10.1088/0004-6256/137/5/4377}

\bibitem[{{Yoon} {et~al.}(2016){Yoon}, {Beers}, {Placco}, {Rasmussen}, {Carollo}, {He}, {Hansen}, {Roederer}, \& {Zeanah}}]{Yoon2016}
{Yoon}, J., {Beers}, T.~C., {Placco}, V.~M., {et~al.} 2016, \apj, 833, 20, \dodoi{10.3847/0004-637X/833/1/20}

\bibitem[{{York} {et~al.}(2000){York}, {Adelman}, {Anderson}, {Anderson}, {Annis}, {Bahcall}, {Bakken}, {Barkhouser}, {Bastian}, {Berman}, {Boroski}, {Bracker}, {Briegel}, {Briggs}, {Brinkmann}, {Brunner}, {Burles}, {Carey}, {Carr}, {Castander}, {Chen}, {Colestock}, {Connolly}, {Crocker}, {Csabai}, {Czarapata}, {Davis}, {Doi}, {Dombeck}, {Eisenstein}, {Ellman}, {Elms}, {Evans}, {Fan}, {Federwitz}, {Fiscelli}, {Friedman}, {Frieman}, {Fukugita}, {Gillespie}, {Gunn}, {Gurbani}, {de Haas}, {Haldeman}, {Harris}, {Hayes}, {Heckman}, {Hennessy}, {Hindsley}, {Holm}, {Holmgren}, {Huang}, {Hull}, {Husby}, {Ichikawa}, {Ichikawa}, {Ivezi{\'c}}, {Kent}, {Kim}, {Kinney}, {Klaene}, {Kleinman}, {Kleinman}, {Knapp}, {Korienek}, {Kron}, {Kunszt}, {Lamb}, {Lee}, {Leger}, {Limmongkol}, {Lindenmeyer}, {Long}, {Loomis}, {Loveday}, {Lucinio}, {Lupton}, {MacKinnon}, {Mannery}, {Mantsch}, {Margon}, {McGehee}, {McKay}, {Meiksin}, {Merelli}, {Monet}, {Munn}, {Narayanan}, {Nash}, {Neilsen}, {Neswold}, {Newberg}, {Nichol}, {Nicinski},
  {Nonino}, {Okada}, {Okamura}, {Ostriker}, {Owen}, {Pauls}, {Peoples}, {Peterson}, {Petravick}, {Pier}, {Pope}, {Pordes}, {Prosapio}, {Rechenmacher}, {Quinn}, {Richards}, {Richmond}, {Rivetta}, {Rockosi}, {Ruthmansdorfer}, {Sandford}, {Schlegel}, {Schneider}, {Sekiguchi}, {Sergey}, {Shimasaku}, {Siegmund}, {Smee}, {Smith}, {Snedden}, {Stone}, {Stoughton}, {Strauss}, {Stubbs}, {SubbaRao}, {Szalay}, {Szapudi}, {Szokoly}, {Thakar}, {Tremonti}, {Tucker}, {Uomoto}, {Vanden Berk}, {Vogeley}, {Waddell}, {Wang}, {Watanabe}, {Weinberg}, {Yanny}, {Yasuda}, \& {SDSS Collaboration}}]{York2000}
{York}, D.~G., {Adelman}, J., {Anderson}, John~E., J., {et~al.} 2000, \aj, 120, 1579, \dodoi{10.1086/301513}

\bibitem[{{Yuan} {et~al.}(2015){Yuan}, {Liu}, {Xiang}, {Huang}, \& {Chen}}]{Yuan2015}
{Yuan}, H., {Liu}, X., {Xiang}, M., {Huang}, Y., \& {Chen}, B. 2015, \apj, 803, 13, \dodoi{10.1088/0004-637X/803/1/13}

\bibitem[{{Zepeda} {et~al.}(2023){Zepeda}, {Beers}, {Placco}, {Shank}, {Gudin}, {Hirai}, {Mardini}, {Pifer}, {Catapano}, \& {Calagna}}]{zepeda2023}
{Zepeda}, J., {Beers}, T.~C., {Placco}, V.~M., {et~al.} 2023, \apj, 947, 23, \dodoi{10.3847/1538-4357/acbbcc}

\bibitem[{{Zhao} {et~al.}(2012){Zhao}, {Zhao}, {Chu}, {Jing}, \& {Deng}}]{Zhao2012}
{Zhao}, G., {Zhao}, Y.-H., {Chu}, Y.-Q., {Jing}, Y.-P., \& {Deng}, L.-C. 2012, Research in Astronomy and Astrophysics, 12, 723, \dodoi{10.1088/1674-4527/12/7/002}

\bibitem[{{Zheng} {et~al.}(2018){Zheng}, {Zhao}, {Wang}, {Fan}, {Tan}, {Li}, \& {Zuo}}]{Zheng2018}
{Zheng}, J., {Zhao}, G., {Wang}, W., {et~al.} 2018, Research in Astronomy and Astrophysics, 18, 147, \dodoi{10.1088/1674-4527/18/12/147}

\end{thebibliography}
